\documentclass{aa}  

\usepackage{natbib}
\bibpunct{(}{)}{;}{a}{}{,} 
\usepackage{graphicx}
\usepackage{txfonts}
\usepackage{twoopt}
\usepackage[breaklinks=true,colorlinks=true,allcolors=blue]{hyperref}
\usepackage{pifont}
\usepackage{array}
\usepackage{adjustbox}
\usepackage{enumitem}
\usepackage{xcolor}
\usepackage{placeins}
\usepackage{multirow}

\makeatletter
  \newcommandtwoopt{\citeads}[3][][]{\href{http://adsabs.harvard.edu/abs/#3}%
    {\def\hyper@linkstart##1##2{}%
     \let\hyper@linkend\@empty\citealp[#1][#2]{#3}}}
  \newcommandtwoopt{\citepads}[3][][]{\href{http://adsabs.harvard.edu/abs/#3}%
    {\def\hyper@linkstart##1##2{}%
     \let\hyper@linkend\@empty\citep[#1][#2]{#3}}}
  \newcommandtwoopt{\citetads}[3][][]{\href{http://adsabs.harvard.edu/abs/#3}%
    {\def\hyper@linkstart##1##2{}%
     \let\hyper@linkend\@empty\citet[#1][#2]{#3}}}
  \newcommandtwoopt{\citeyearads}[3][][]%
    {\href{http://adsabs.harvard.edu/abs/#3}
    {\def\hyper@linkstart##1##2{}%
     \let\hyper@linkend\@empty\citeyear[#1][#2]{#3}}}
\makeatother

\def\hi{H\,{\sc i}}
\newcommand\kms{km~s$^{-1}$}
\newcommand\msun{$M_\odot$}
\newcommand\mhi{$M_{\text{\hi}}$}
\def\arcmin{$^{\prime}$}
\def\arcsec{$^{\prime\prime}$}
\def\dg{$^{\circ}$}
\def\w50{$W_{50}$}

\def\mstar{$M_{\star}$}
\newcommand {\um}{\,{$\mu$\rm m}}
\newcommand {\Mpc} {\,{\rm Mpc}}

\newcommand {\kmsMpc} {\,{\rm km\,s}^{-1}\,{\rm \Mpc}^{-1}}
\def\be{\begin{equation}}
\def\ee{\end{equation}}

\def\vsys{$V_{\rm sys}$}

\def\mbar{$M_{\rm b}$}

\def\smrBesla{$2\times 10^8$ < \mstar / \msun\ < $5\times 10^9$}
\def\mstarmassive{$M_{\star}^{\rm massive}$}

\def\deltavsys{|$\Delta$\vsys|}

\def\dSFR{$\Delta \log$ SFR}
\def\mvir{$M_{\rm 200}$}
\def\Rvirpair{$R^{\rm pair}_{\rm 200}$}
\def\rp{$r_{\rm p}$}

\usepackage{subcaption}
\begin{document}

   \title{Gas-rich dwarf galaxy multiples in the Apertif \hi\ survey}

   \subtitle{}

   \author{
   B. Šiljeg\inst{1,2}
   \and
   E. A. K. Adams\inst{1,2}
   \and
   F. Fraternali\inst{2}
   \and
   K. M. Hess\inst{3,1}
   \and
   A. Marasco\inst{4}
   \and
   H. Dénes\inst{5,1}
   \and
   J. Garrido\inst{6}
   \and
   D. M. Lucero\inst{7}
   \and
   R. Morganti\inst{1,2}
   \and
   S. Sánchez-Expósito\inst{6}
   \and
   J. M. van der Hulst\inst{2}
   }

    \institute{
    ASTRON, the Netherlands Institute for Radio Astronomy, Oude Hoogeveenseweg 4, 7991 PD Dwingeloo, The Netherlands
    \and
    Kapteyn Astronomical Institute, University of Groningen, P.O. Box 800, 9700 AV, Groningen, The Netherlands
    \and
    Department of Space, Earth and Environment, Chalmers University of Technology, Onsala Space Observatory, 43992 Onsala, Sweden
    \and
    INAF - Padova Astronomical Observatory, Vicolo dell’Osservatorio 5, I-35122 Padova, Italy
    \and
    School of Physical Sciences and Nanotechnology, Yachay Tech University, Hacienda San José S/N, 100119, Urcuquí, Ecuador
    \and
    Instituto de Astrof\'{i}sica de Andaluc\'{i}a (CSIC), Glorieta de la Astronom\'{i}a s/n, 18008 Granada, Spain
    \and
    Department of Physics, Virginia Polytechnic and State University, Blacksburg, VA 24061-0435, USA
    }

   \date{}

  \abstract
   {Dwarf-dwarf galaxy encounters are a key aspect of galaxy evolution as they can ignite or temporarily suppress star formation in dwarfs and can lead to dwarf mergers. However, the frequency and impact of dwarf encounters remain poorly constrained due to limitations of spectroscopic studies, e.g. surface-brightness incompleteness of optical studies and poor spatial resolution of single-dish neutral hydrogen (\hi) surveys.}
   {We aim to quantify the frequency of isolated gas-rich dwarf galaxy multiples using the untargeted, interferometric Apertif \hi\ survey and study the impact of the interaction on star formation rates of galaxies as a function of the on-sky separation.}
   {Our parent dwarf sample consists of 2481 gas-rich galaxies with stellar masses $ \sim 10^6 < M_{\star} / M_{\odot} \lesssim 5 \times 10^9$, for which we identify close companions based on projected separation (\rp) and systemic velocity difference (\deltavsys). We explore both constant thresholds for \rp\ and \deltavsys\ corresponding to 150 kpc and 150 \kms\ on all galaxies in our sample, and mass-dependent thresholds based on a stellar-to-halo mass relation.}
   {We find the average number of companions per dwarf in our sample to be 13\% (20\%) 
   when considering mass-dependent (constant) thresholds.
   In the stellar mass regime of  $2\times 10^8 < M_{\star} / M_{\odot} < 5 \times 10^9$, we find a three times higher frequency ($\sim$11.6\%) of dwarf companions than previously determined from optical spectroscopic studies, highlighting the power of \hi\ for finding dwarf multiples. Furthermore, we find evidence for an increase in star formation rates (SFRs) of close dwarf galaxy pairs of galaxies with similar stellar masses.}
   {}


   \maketitle

\section{Introduction}
\label{p1:sec:introduction}

    Galaxy interactions play a crucial role in galaxy evolution by, for example, enhancing star-formation \citepads[e.g.][]{2008MNRAS.385.1903L,2008AJ....135.1877E}, tidally disturbing morphologies of galaxies, and redistributing their gas reservoirs \citepads[e.g.][]{1992ARA&A..30..705B,2000RSPTA.358.2063S,2016MNRAS.459..720H}.
    Furthermore, they can lead to mergers which are thought to also trigger active galactic nuclei (AGN) \citepads[e.g.][]{2014MNRAS.441.1297S,2018PASJ...70S..37G,2020A&A...637A..94G,2024A&A...690A.326L,2025A&A...699A.330E} and to be the main drivers of morphological transformations from discs to spheroidal systems \citepads[e.g.][]{2018MNRAS.480.2266M}. However, even non-disruptive flyby events can significantly impact properties of galaxies and affect their later secular evolution \citepads[e.g.][]{1996Natur.379..613M,2021MNRAS.500.4937M,2021MNRAS.506...98K}. Hence, determining the frequency of such galaxy encounters and their impact on galaxies across all mass scales provides a direct constraint on galaxy evolution models.

    Most observational and theoretical studies have been focused on massive galaxies, both for studying the pair fractions \citepads[e.g.][]{1995ApJ...445...37Y,2007ApJS..172..320K,2019ApJ...876..110D} and the impact of interactions on galaxy properties \citepads[e.g.][]{2006MNRAS.367.1029S,2008AJ....135.1877E,2022MNRAS.513.1867D}. However, due to the low potential wells of dwarf galaxies (galaxies with stellar masses \mstar\ < $5\times 10^9$ \msun), interactions likely play an even greater role in their evolutionary pathway as they are more easily disturbed and take longer to resettle to equilibrium \citepads[e.g.][]{2021MNRAS.500.4937M}. Yet, dwarf-dwarf encounters remain poorly constrained both in their frequency and influence on galaxy properties. Previous studies have already hinted toward a difference between the impact galaxy interactions have on dwarf galaxies when compared to high mass galaxies, showing that interactions can both enhance and temporarily suppress star formation rates (SFRs) of dwarf galaxies \citepads{2015ApJ...805....2S,2024ApJ...963...37K,2025ApJ...980..157H}, while only enhancement has been observed in higher mass systems \citepads[e.g.][]{2006MNRAS.367.1029S,2025ApJ...980..157H}. In fact, \citetads{2025ApJ...980..157H} suggested that the SFR suppression and enhancement in dwarfs in gas-rich pairs (including both dwarf-dwarf and dwarf-high mass pairs) depends on the proximity of interacting galaxies, pointing toward a distinct dominant mechanism in interacting dwarf galaxies which is likely linked to high gas-fractions and low potential wells of dwarfs when compared to high-mass galaxies.

    Previous studies on the frequency of dwarf galaxy encounters are primarily based on optical spectroscopic studies with large sky coverage. \citetads{2013MNRAS.428..573S} investigated how the average number of satellite galaxies varies with the stellar mass of the host using data from the Sloan Digital Sky Survey (SDSS). For dwarf host masses (\mstar\ $\lesssim 10^{10}$\msun), they find that the average number of satellites becomes independent of the host mass, and is of the order of a few percent for stellar mass ratios above 1/10. Similarly, \citetads{2018MNRAS.480.3376B} (hereafter B18) studied the frequency of dwarf galaxy multiples found in the SDSS catalog and reported the fraction of $\sim$4\% of dwarfs having a close-by companion within the stellar mass range of $(2\times 10^8 - 5\times 10^9)$ \msun. By comparing to the \textit{Illustris-1} cosmological simulation \citepads{2014Natur.509..177V,2015A&C....13...12N}, they predict this fraction to increase up to $\sim$6\% in future surveys with improved completeness at these masses. Generally, the frequency of dwarf encounters seems extremely rare, that is, of the order of a few percent.
    
    However, optical spectroscopic studies are inherently limited by fiber collisions \citepads{2002AJ....124.1810S} which limit the detection of close galaxy pairs, and surface brightness limitations which preferentially exclude diffuse, low surface brightness dwarfs \citepads[e.g.][]{2005ApJ...631..208B}. Galaxy interactions can also significantly change morphologies of dwarf galaxies and potentially lower their surface brightness \citepads[e.g.][]{2018ApJ...866L..11B,2019MNRAS.488.2743T}. On the other hand, by enhancing the SFR, interactions can also increase dwarf's surface brightness. 
    Depending on the relative contribution of these two effects in a dwarf-dwarf interaction, optical selection could potentially be biased for or against the detection of the interacting pair when compared to single dwarfs. Hence, due to technical limitations and the aforementioned biases, it is hard to robustly constrain the true number of dwarf galaxy encounters using current optical spectroscopic surveys.
    
    Neutral hydrogen (\hi) observations provide an alternative method for finding dwarf galaxies. Dwarfs in the field (isolated from high-mass galaxies) tend to be gas-rich with increasing gas fractions toward lower stellar masses \citepads[e.g.][]{2006ApJ...653..240G,2012ApJ...756..113H,2018MNRAS.476..875C}. In fact, \hi\ is often the dominant baryonic component of these systems, making \hi\ observations an extremely useful tool for finding dwarf galaxies. The Arecibo Legacy Fast ALFA (ALFALFA) \hi\ survey \citepads{2005AJ....130.2598G}, a single dish untargeted \hi\ survey conducted with the Arecibo telescope, has already shown how effective \hi\ observations can be in identifying galaxies missed by optical catalogs \citepads[e.g.][]{2015AJ....149...72C,2015ApJ...801...96J,2017ApJ...842..133L}. Hence, \hi\ surveys can provide a field dwarf galaxy sample that spans the whole range of dwarf diversity, independently of their stellar content. 
    
    Previous large-scale untargeted \hi\ surveys (e.g. HIPASS, \citeads{2004MNRAS.350.1195M}; ALFALFA, \citeads{2005AJ....130.2598G,2018ApJ...861...49H}) relied on single-dish telescopes which lacked the spatial resolution needed to resolve individual dwarf galaxies in close pairs (e.g. 3.5\arcmin\ for Arecibo, corresponding to $\sim$70 kpc at 70 Mpc distance; \citeads{2005AJ....130.2598G}). The recent onset of interferometric, untargeted \hi\ surveys (e.g. Apertif, \citeads{2022A&A...667A..38A}; MIGHTEE, \citeads{jarvis2017meerkatinternationalghztiered}; WALLABY, \citeads{2020Ap&SS.365..118K}) are now enabling us to both detect and resolve dwarf galaxy multiples as well as isolated dwarfs, making these surveys an extremely powerful tool for constraining the true frequency of dwarf-dwarf encounters in observations.
    
    In this work, we use $\sim$42\% of the total spatial coverage and $\sim$33\% of the total spectral coverage of the Apertif \hi\ survey (\citeads{2022A&A...667A..38A}; Hess et al., in prep.) to constrain the frequency of dwarf encounters and explore the properties of paired galaxies. Our parent sample contains 2481 gas-rich candidate dwarf galaxies within the stellar mass range of $\sim 10^6 - 5\times 10^9$ \msun. We identify pairs and multiples using projected separation and the difference in their systemic velocities, and compare our findings with previous work based on optical spectroscopy. Furthermore, we explore the SFR dependence on the projected separation of paired dwarfs and compare with previous work on gas-rich paired galaxies.

    The paper is structured as follows: In Section \ref{p1:sec:data}, we describe the Apertif \hi\ survey and measurements of properties of Apertif detections. Section \ref{p1:sec:dwarf_sample} describes the selection of the parent dwarf sample and gives an overview of its properties. In Section \ref{p1:sec:analysis}, we present our methodology for identifying dwarf multiples and the isolation criteria. In Section \ref{p1:sec:results}, we report the frequency of dwarf multiples found in Apertif and explore the effects of interactions on the star formation activity of paired dwarfs. In Section \ref{p1:sec:discussion}, we compare the frequency of dwarf multiples with an optical study by B18, and the SFRs of paired dwarfs with previous study by \citetads{2025ApJ...980..157H}. 
    Finally, in Section \ref{p1:sec:conclusions} we draw our conclusions.

\section{Data}
\label{p1:sec:data}

\subsection{Apertif \hi\ data}
\label{p1:sec:apertif_data}

    Apertif \citepads{2022A&A...658A.146V} is a phased-array feed receiver system designed for the Westerbork Synthesis Radio Telescope (WSRT). It increased the field of view of the telescope up to 8 deg$^2$ by producing 40 instantaneous compound beams on the sky. This made WSRT (together with Apertif) a great instrument for conducting a wide area survey across the northern sky.

    The Apertif imaging survey \citepads{2022A&A...667A..38A} observed selected areas of the sky with declination above 25\dg. Each Apertif observation is 11.5 hours, with simultaneous execution of neutral hydrogen (\hi) spectral line, continuum and polarization imaging. 
    The \hi\ cubes are produced over the topocentric frequency range 1292.5 - 1429.3 MHz \citepads{2022A&A...667A..38A} and have spatial resolution of $ 15$\arcsec$ \times 15$\arcsec$ / \sin \delta $. The Apertif imaging surveys are split in two tiers: the Apertif Wide-area Extragalactic Survey (AWES) that covers $\sim$ 2200 deg$^2$ of the sky with one or two observations per field; and the Medium-Deep Survey (MDS) that targets specific areas of the sky with $\sim 130$ deg$^2$ of coverage and up to 12 observations per field \citepads[see][]{2022A&A...667A..38A}.

    Spectral line cubes for Apertif data were split into four subsets in spectral dimension for easier data handling. In this work, we focus on a subset which covers line-of-sight velocities from 424 to 10,170 \kms (cube 2; see \citeads{2022A&A...667A..38A} for a full description of the cube properties). These cubes were the initial targets for source finding as they provide maximal number of detections by optimizing sensitivity over the velocity channel width of 7.7 \kms\ (Hess et al., in prep.). We use a subset of this data that covers $\sim$42\% of Apertif spatial footprint, as the source finding on the remaining footprint is still ongoing.

    \subsection{Apertif \hi\ processing}
    \label{p1:sec:Apertif_HI_processing}
    
    The \hi\ post-processing workflow is publicly available\footnote{\url{https://github.com/kmhess/aper_sf2/}}, 
    and will be described in a future paper. Here, we provide a brief overview.
    
    Within the Apertif imaging survey footprint, observations of individual fields have been co-added with channel based weights (when multiple observations were available), and the 40 compound beams within individual fields have been mosaicked. This ensured the best possible S/N per field and fairly uniform sensitivity within individual fields, thereby maximizing the probability of detecting low \hi\ mass sources. However, the obtained sensitivity varies on larger scales ($\gtrsim$ 2\dg; larger than the size of an individual field) due to differing numbers of observations per field and instrumental effects (e.g., continuum subtraction) within the Apertif cubes themselves. As a result of the former, mosaicking of neighboring fields has also not yet been optimized for spectral line source finding. On average, the rms noise level around detected sources corresponds to $\sim 1.3_{-0.3}^{+0.6}$ mJy beam$^{-1}$.

    Spectral line source finding (see Section \ref{p1:sec:HI_source_finding}) was performed on the mosaicked data. The candidate sources were vetted to remove continuum artefacts and sidelobes from bright \hi\ sources before cleaning. Cleaning was then performed in the image plane, on a per-compound beam basis, using the Apercal \citepads{2022A&C....3800514A} implementation of Miriad's \texttt{clean} and \texttt{restor} tasks \citepads{2011ascl.soft06007S}. After cleaning, the cleaned compound beams were re-mosaicked, and the mask was applied to the cleaned mosaic to produce advanced data products using the SoFiA Image Pipeline (SIP)\footnote{\url{https://github.com/kmhess/SoFiA-image-pipeline}} \citepads{kelley_m_hess_2024_14016958}.

\subsection{The \hi\ source finding}
\label{p1:sec:HI_source_finding}

    The \hi\ source finding on Apertif cubes (Hess et al., in prep.) was conducted on individual Apertif fields using the SoFiA-2 source finder\footnote{\url{https://gitlab.com/SoFiA-Admin/SoFiA-2}} \citepads{2015MNRAS.448.1922S,2021MNRAS.506.3962W}. 
    SoFiA-2 is run using three spatial kernels with FWHMs of: 0, 3, and 6 pixels (corresponding to spatial resolutions of 15\arcsec, 23.4\arcsec, and 39\arcsec); and three boxcar spectral kernels of sizes: 0, 3, and 7 channels (corresponding to spectral resolutions of 7.7 \kms, 23.2 \kms, and 54.1 \kms; Hess et al., in prep.). The mask of each candidate detection is then constructed from pixels that have $S/N$ > 3.8 (threshold obtained by the optimization of true vs. false positive detections) in one or more combinations of spatial and spectral smoothing kernels. The mask is constructed for both positive and negative candidates, whose properties are afterwards compared and used to correct for false positives. We used a reliability threshold of 0.65 for this comparison to improve the completeness. Candidates that passed the test and subsequent vetting (see Sect. \ref{p1:sec:Apertif_HI_processing}) were then cleaned. After cleaning and mosaicking, the pipeline uses SIP to produce inspection plots containing a Pan-STARRS 1 image \citepads{2016arXiv161205560C} overlaid with \hi\ contours, a total \hi\ intensity map, an $S/N$ map, a velocity map, a position-velocity (PV) slice, and a global spectral profile. These inspection plots are made at the original and 40\arcsec\ resolution for each source, and both are visually inspected before a source is included in the final source list.

\subsection{Distance and \hi\ masses of detected sources}
\label{p1:sec:distance_and_HI_mass}

    We calculate line-of-sight velocity-based distances using the linear density field model of \citetads{2024NatAs...8.1610V} provided by the Extragalactic Distance Database (EDD) \citepads{2009AJ....138..323T,2020AJ....159...67K}. 
    This model is calibrated on the Cosmicflows-4 (CF4) database of galactic distances \citepads{2023ApJ...944...94T}, a compendium of redshift-independent distance measurements for $\sim\!56,000$ nearby ($D\!\lesssim500\ \!\Mpc$) galaxies. We adopted the error estimation procedure from \citetads{2025A&A...696A.185H}, corresponding to roughly 12\% error on average for the Apertif source list. 
    Flow model-based distances for Apertif detections differ from Hubble distances (assuming $H_0\!=\!75\kmsMpc$, which is compatible with the CF4 measurements) by $7\%$ on average, and up to $31\%$.

    We estimate \hi\ masses of each detection using the following relation \citepads[e.g.][]{2012ARA&A..50..531K}:
    \begin{equation}
        \frac{M_{\text{\hi}}}{\rm{M_{\odot}}} = 2.343\cdot 10^5\cdot \left(\frac{D}{\rm Mpc}\right)^2 \cdot \frac{F_{\text{\hi}}}{\rm Jy\, km/s}
        \label{p1:eq:HI_mass}
    \end{equation}
    where $F_{\text{\hi}}$ is the total \hi\ flux, and $D$ the distance to the galaxy. We assume a 15\% error on $F_{\text{\hi}}$ coming from the calibration of the flux scale, primary beam correction and mosaicking of Apertif compound beams \citepads{2022A&A...667A..39K}.

\subsection{Stellar properties of Apertif detections}
\label{p1:sec:stellar_properties}
    
    Stellar masses ($M_\star$) and SFRs of galaxies associated with our \hi\ detections are determined following the approach of \citetads{2019ApJS..244...24L}, based on a combination of near-infrared (NIR), mid-infrared (MIR), near-ultraviolet (NUV) and far-ultraviolet (FUV) photometry measurements.
    As in \citetads[][hereafter M23]{2023A&A...670A..92M}, we focus on archival, publicly available images from the Wide-field Infrared Survey Explorer \citepads[WISE;][]{2010AJ....140.1868W} in bands W1 ($3.4$\um) and W4 ($22$\um), from the IRAC $3.6$\um\ and MIPS $24$\um\ camera onboard of the \emph{Spitzer Space Telescope} \citepads{2004ApJS..154....1W}, and from the Galaxy Evolution Explorer \citepads[\emph{GALEX;}][]{2005ApJ...619L...1M} for both UV bands.
    We briefly summarize our procedure below, and redirect the reader to Section 2 and Appendix A of M23 for further details.
    
    Counterpart identification is done galaxy by galaxy by visual inspection of the moment-0 \hi\ maps overlaid on top of the UV images. 
    For systems that are too faint in UV or that are not covered by the \emph{GALEX} footprint, we look for non point-like objects in W1 (or in the IRAC $3.6$\um\ band, if available) that have a counterpart in W4.
    We stress that, for low-mass or distant galaxies, the counterpart identification is particularly problematic and the chance of selecting a foreground star is not zero.   
    We dealt with this issue a posteriori by cross-matching the Apertif source list with the stellar catalog from the Gaia DR3 \citepads{2023A&A...674A...1G}, and visually inspecting all detections whose photometric centroid is closer than the FWHM of W1 (6.1\arcsec) from a Gaia detection. With this, we excluded 138 out of 413 inspected cases. In addition, we visually inspected all the remaining detections for which the measured half-light radius is smaller than 3\arcsec\ (half of the FWHM of the W1 PSF), and have excluded 23 more detections (out of 58 inspected cases).
    Thus, 161 detections (out of 471 visually inspected cases) have been flagged as having unreliable stellar measurements in the Apertif source list of 3943 detections. These measurements will not be taken into account in the final construction of the parent dwarf galaxy sample (see Sect. \ref{p1:sec:dwarf_sample}).  

    Photometry for the identified sources is done on sky-subtracted images after the automatic removal of contamination from point-like sources such as foreground stars and high-redshift background galaxies. 
    With respect to the original procedure from M23, several steps have been automated following approaches from previous software packages, that is, determination of the inclination and position angles following SExtractor \citepads{1996A&AS..117..393B} and cleaning of the sky region following MORPHOT \citepads{2012MNRAS.420..926F}. In short, the 'cleaned' image is obtained by comparing the original image to the one rotated by 180\dg, finding the pixels for which the residual between the two images is larger than $3\times$ the rms intensity in the surrounding region of the rotated image, and replacing these pixels values with the ones from the rotated image.
    Full details on these improvements are provided in Appendix A of \citetads{2025A&A...695L..23M}.
    Additionally, we have developed an interactive tool to manually mask bright, extended objects surrounding the target.
    This tool was applied to the vast majority of images, especially in the W1 band, to minimise the contamination from surrounding systems.
    Photometric measurements are then extracted as cumulative light profiles. We note that our photometric analysis does not systematically ensure the flattening of the cumulative light profile has been reached. However, we later tested the flattening of the profiles of galaxies that entered our final sample, proving the flattening has indeed been reached for the vast majority galaxies.

    The $M_\star$ and SFR derivation from \citetads{2019ApJS..244...24L} is calibrated on the results of population synthesis modelling of the \emph{GALEX}-SDSS-WISE Legacy Catalogue by \citetads{2016ApJS..227....2S,2018ApJ...859...11S}. In short, UV and MIR data are used to determine the SFR of a galaxy: this is employed as a color-correction term for the $M_\star$-to-$L_{\rm NIR}$ ratio (NIR being IRAC 3.6\um\ if available, or W1 otherwise), that in turn is used to infer $M_\star$.
    Typical values for our sources correspond to mass-to-light ratios in the NIR band of $0.35^{+0.14}_{-0.07}$.
    Although the procedure of \citetads{2019ApJS..244...24L} makes use of IR data from the WISE W1 and W4 bands alone, we prefer to use the higher quality \emph{Spitzer} images from IRAC and MIPS whenever possible. 
    We have verified that, in this way, the scatter in the SFR-$M_\star$ plane is reduced.
    We only use photometric measurements with integrated S/N ratio larger than $3$. Uncertainties on stellar masses are obtained from the quadratic sum of the photometric error (defined as in Equation A.1. of \citeads{2025A&A...695L..23M}), and the error on the galaxy distance.

\subsection{Merged, double, and split detections}
\label{p1:sec:merged_double_detec}

    The Apertif source list used in this work (corresponding to $\sim$42\% of spatial and $\sim$33\% of spectral coverage of Apertif) contains 3943 sources. However, this does not correspond to the true number of \hi\ sources in this footprint due to the presence of merged, double and split detections, as explained below.

    During the source finding, emission from two (or more) galaxies can be mistakenly selected as a single \hi\ source if they are close-by on sky and in line-of-sight velocity space. We refer to these cases as merged detections. The conditions on when the merging happens depend on the geometry of individual systems. 
    As mentioned in Sect. \ref{p1:sec:HI_source_finding}, multiple resolution kernels were used when running SoFiA on Apertif data (from 15\arcsec\ to 39\arcsec\ spatially and from 7.7 \kms\ to 54.1 \kms\ spectrally) to search for detections. 
    Once the source masks are constructed using all combinations of spatial and spectral kernels, two separate sources can still be merged into one if their masks are within 2 pixels (12\arcsec) and 3 channels (23.2 \kms) of each other.
    
    On the other hand, as the source finding was run on individual Apertif fields whose footprints overlap, the source list can contain several detections of the same source. We call these occurrences double detections. In addition, emission from one source can be split in 2 detections for low signal-to-noise sources, when there is a flagged channel at velocities corresponding to the galaxy emission, or in cases of massive and highly inclined galaxies where approaching and receding sides get separated due to low S/N of central channels
    (for more details see Hess et al., in prep.). The double and split detections can together create multiple detections of the same source. 

    With the caveats described above, we note that the Apertif source list is not representative of the true number of galaxies detected in Apertif. Therefore, for a statistical study such as this, we first need to clean the source list (find and correct for merged, double, and split detections) at dwarf scales, thereby constructing the final dwarf galaxy sample.

\section{Apertif dwarf galaxy sample}
\label{p1:sec:dwarf_sample}

    We select dwarf galaxy candidates from the Apertif source list by applying the upper stellar mass limit of $5\times 10^{9}$ \msun. We also include higher stellar mass galaxies for which $5\times 10^{9}$ \msun\ is within the 1$\sigma$ error of the stellar mass measurement. 
    This selection led to our initial sample of 2795 sources. However, as mentioned in Sect. \ref{p1:sec:merged_double_detec}, \hi\ detections in the Apertif source list do not always correspond to the emission from one galaxy. 
    Hence, in order to constrain the final number of detected dwarf galaxies, we need to correct for merged and multiple detections, which we describe in the following section. The main steps of the procedure are illustrated in Fig. \ref{p1:fig:flowchart}.

\subsection{Correction for merged and multiple detections}
\label{p1:sec:producing_sample}

    \begin{figure*}
        \centering
        \includegraphics[width = 1 \textwidth]{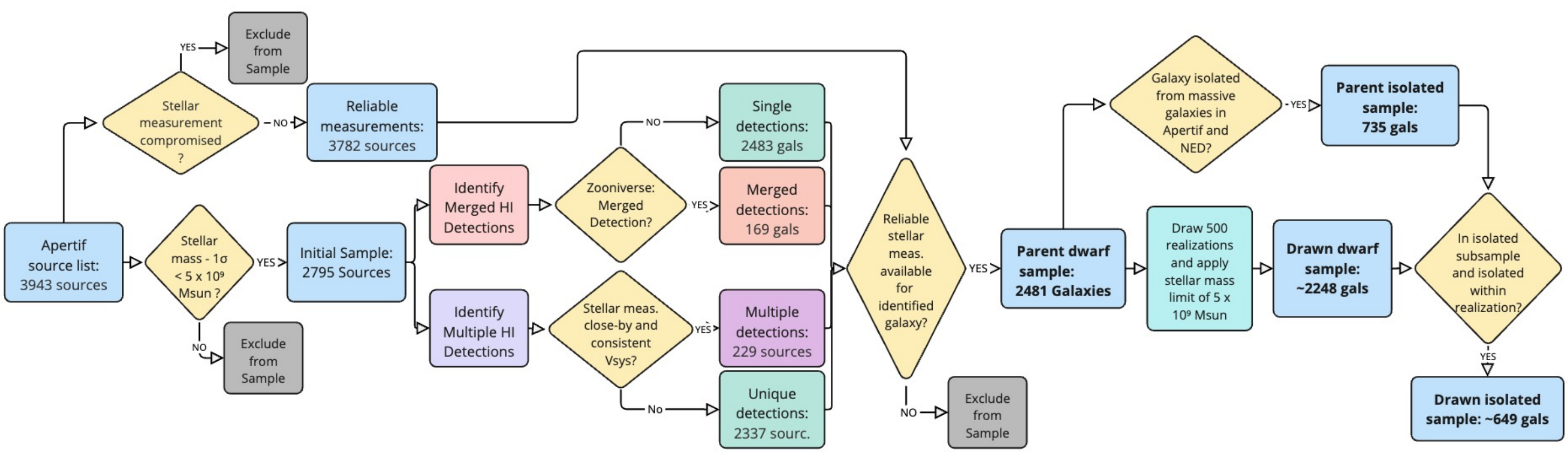}\\
        \caption{Selection of dwarf galaxy samples from Apertif. The flowchart denotes most relevant steps in our selection procedure starting from the Apertif source list to dwarf galaxy samples used in this work. Boxes with bold text denote final samples that we use in our analysis.}
        \label{p1:fig:flowchart}
    \end{figure*}
    
    To identify merged detections, we make use of Zooniverse\footnote{https://www.zooniverse.org/}, by constructing a private project with all 2795 dwarf candidate sources in our initial sample. We use optical images from PanSTARRS-1 photometric survey, infrared images from WISE, and ultraviolet images from GALEX, all overlaid with \hi\ contours to visually classify each detection as corresponding to one or multiple galaxies. In some cases, we have additionally inspected SIP outputs for Apertif \hi\ detections to confirm our classification. With this, we identify 82 merged detections. For a subset of 11, the \hi\ is poorly resolved, low S/N, or both, making it unclear if there are indeed multiple optical galaxies associated with the \hi\ or a happenstance alignment of multiple optical galaxies at different distances. For two of these cases, the multiple was confirmed by optical redshifts of candidate galaxies obtained from the NASA/IPAC Extragalactic Database\footnote{\url{https://ned.ipac.caltech.edu/}} (NED), while for the other nine cases there were no optical redshift measurements available for the confirmation. 
    These nine cases have been classified as 'unclear' merged detections; see Appendix \ref{p1:app:unclear_merged} for more information on these cases. In the rest of the paper, we will treat them as merged detections, noting that results do not change significantly by excluding these sources from the sample. For each merged detection, we used the same methodology described in Sect. \ref{p1:sec:stellar_properties} to measure stellar properties of companion galaxies. 
    As we do for detections of single sources, we adopt individual galaxies in our sample only if their stellar properties could be reliably measured (e.g. no strong contamination by foreground stars). We apply this criteria independently on the companion and the original (previously measured) galaxy. 

    We searched for multiple detections of the same source based on the coordinates of the photometric measurements. We visually inspect all detections that have photometric measurements closer than 35\arcsec\ on the sky. This relatively large radius is chosen because differences in the \hi\ detections (especially split detections that can be spatially offset on the sky) can lead to the selection of different optical counterparts. Increasing this radius to 45\arcsec\ ($\sim$ 3 Apertif beams) did not provide any new multiple detection candidates. We then also checked for consistent recessional velocities between the duplicate detection candidates. In the end, 229 (out of 230) sources were confirmed to have been detected multiple times. For each multiple detection, we then chose one measurement of stellar and \hi\ components of corresponding galaxies by visually inspecting \hi\ total intensity maps and positions of available stellar measurements. For split detections, we summed the \hi\ flux of the two detections in order to estimate the true \hi\ mass of the galaxy.

    After correcting for merged detections, multiple detections, and galaxies whose stellar measurement was compromised by a nearby star or a background galaxy (see Sect. \ref{p1:sec:stellar_properties}), we end up with 2481 dwarf candidate galaxies which we refer to as the parent dwarf galaxy sample (Fig. \ref{p1:fig:flowchart}). The \hi\ properties (systemic velocity and the \hi\ mass) of galaxies in merged detections were measured by manually splitting the data cube (spatially) and extracting the spectrum of each galaxy individually. If galaxies in the merged detection were not spatially resolved in \hi\ (25 merged detections; from which 48 galaxies entered the parent dwarf sample), we simply divided the total \hi\ mass of the merged detection by the number of galaxies corresponding to that detection, and assigned to them the same \vsys\ corresponding to the merged detection.

\subsection{Baryonic and dark matter halo masses}
\label{p1:sec:bary_halo_masses}   
    
    For each galaxy in our parent dwarf galaxy sample, we calculate its baryonic mass, using:
    \begin{equation}
        \text{\mbar} = \text{\mstar} + 1.33\text{\mhi}
    \end{equation}
    where a factor of 1.33 corrects for the contribution of helium to the gas mass of galaxies. We neglect the contribution of molecular gas in the calculation of \mbar\ as molecular gas mass in low-mass galaxies has been shown to be significantly smaller than \hi\ and stellar masses (around 10\% of either) \citepads[see e.g.][]{2009AJ....137.4670L,2014MNRAS.445.2599B,2017MNRAS.470.4750A,2018MNRAS.474.4366P,2018MNRAS.476..875C}. 
    
    We assign dark matter halo masses to galaxies in our parent sample using the stellar-to-halo mass relation (SHMR) from the semi-empirical dwarf galaxy formation model DarkLight \citepads{2024arXiv240815214K} applied to the dark-matter-only simulation of the void volume from the Engineering Dwarfs at Galaxy formation’s Edge (EDGE) project. The model is tailored to accurately predict dwarf galaxy properties down to the lowest masses by incorporating the relation between the SFR and dark matter halo properties pre- and post-reionisation. Its application to the void volume provides us with a perfect relation for our \hi\ selected (hence, naturally more isolated) sample of dwarf galaxies. The SHMR relation is:
    \begin{equation}
    \label{p1:eq:SHMR}
        \log \frac{\text{\mvir}}{\text{\msun}} = 0.57 \log \frac{\text{\mstar}}{\text{\msun}}  + 6.08
    \end{equation}
    where \mvir\ is the virial mass of the dark matter halo, defined as the mass within $R_{200}$, the radius within which the mean dark matter density is equal to 200 times the critical density of the Universe. We do not take into account the scatter of the relation as our goal is to trace the mass dependent effects in the selection of dwarf multiples (see Sect. \ref{p1:sec:analysis}), for which introducing the scatter would bring inconsistencies into our analysis (e.g. galaxies of the same stellar mass having different thresholds in projected separation and systemic velocity difference for companion assignment).

\subsection{Properties of the parent dwarf galaxy sample}
\label{p1:sec:sample_properties}

    \begin{figure*}
        \centering
        \includegraphics[width = 0.95 \textwidth]{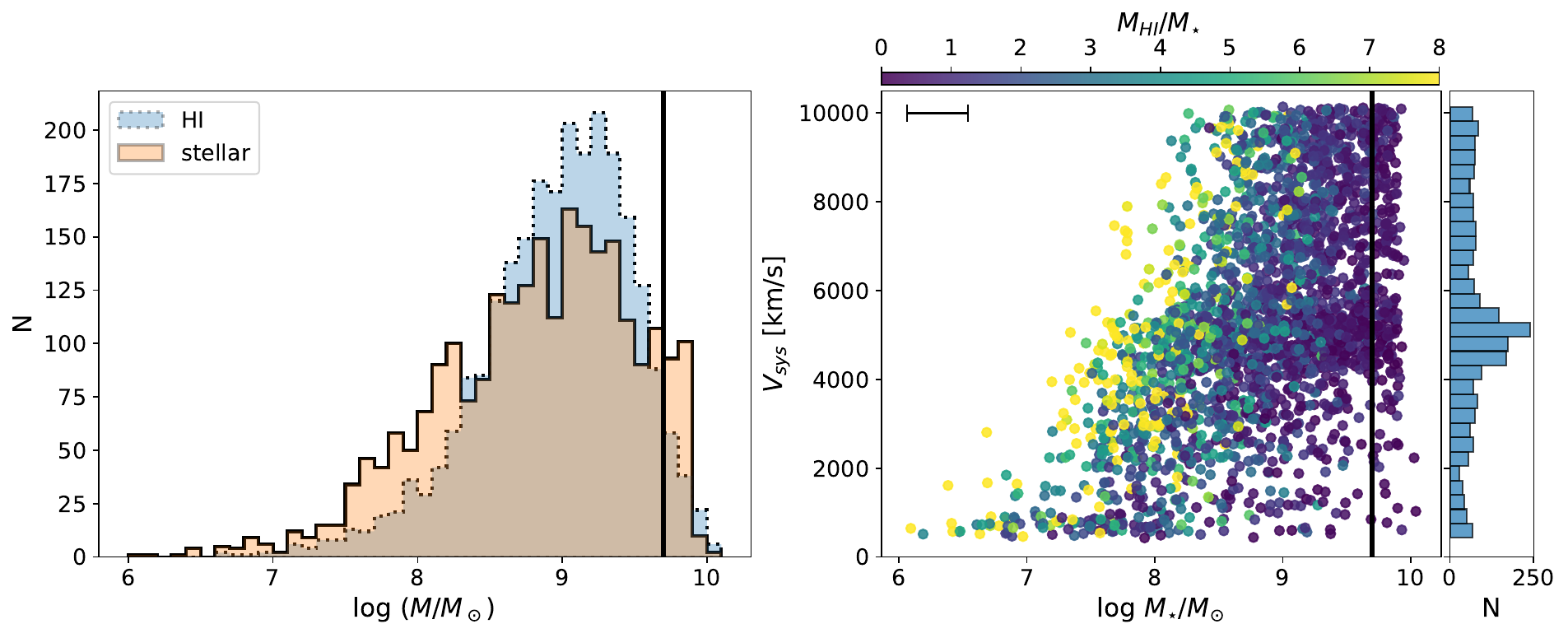}\\
        \caption{Properties of the parent dwarf galaxy sample. \textit{Left:} Histograms of \hi\ (blue) and stellar (orange) masses of the sample. The vertical line denotes the dwarf stellar mass limit of $5\times 10^9$\msun. \textit{Right:} Systemic velocity versus the stellar mass of dwarfs in the sample, color coded by the gas fraction. We show typical stellar mass error in the upper left corner. The vertical line is the same as in the left plot. On the right, we plot the histogram along the \vsys\ axis of our sample.
        }
        \label{p1:fig:sample_properties}
    \end{figure*}

    We show the distributions of stellar and \hi\ masses of our parent dwarf sample in the left panel of Fig. \ref{p1:fig:sample_properties}. As mentioned at the beginning of the section, the sample includes galaxies with \mstar\ measurement within 1$\sigma$ level of our high mass limit of $5 \times 10^9$ \msun. Both distributions peak at masses around 10$^9$ \msun\ and steeply fall towards higher and lower masses. We are able to detect dwarfs down to stellar masses of 10$^6$ \msun, but are biased towards high gas fraction as seen from the excess of stellar mass counts compared to the \hi\ mass counts at the lower masses. This is a natural consequence of our \hi-based selection. A similar excess of stellar mass counts (compared to the \hi-mass ones) is seen also at the highest masses in our sample, which is a consequence of the anti-correlation of gas fraction and stellar mass of \hi\ bearing galaxies \citepads[e.g.][]{2012ApJ...756..113H}. These trends in gas fraction are also seen in the right panel of Fig. \ref{p1:fig:sample_properties} showing the systemic velocity (as a proxy for distance) dependence on the stellar mass of our sample. As expected, the lowest stellar mass galaxies are preferentially detected nearby and are more gas rich. However, for stellar masses around 10$^8$ \msun\ and above, we are able to sample the whole available redshift range of our data. The robust quantification of completeness of our sample, however, is beyond the scope of this work. 

    The histogram of \vsys\ is shown in the right panel of Fig. \ref{p1:fig:sample_properties}. While the distribution is uniform in the majority of our redshift range, there is a peak around systemic velocities of 5000 \kms. 
    This is a signature of the Perseus-Pisces filament within our footprint. In Sect. \ref{p1:sec:isolation_criteria}, we apply a strict isolation criterion to create a subsample of isolated galaxies, in which the overdensity at 5000 \kms\ completely disappears. We then present the properties (Sect. \ref{p1:sec:isolation_criteria}) and pair frequency (Sect. \ref{p1:sec:multiples_Apertif}) of both our full sample and the highly isolated subsample.

\section{Analysis}
\label{p1:sec:analysis}

\subsection{Selection of companions}
\label{p1:sec:selection_companions}

    \begin{figure}
        \centering
        \includegraphics[width = 0.5 \textwidth]{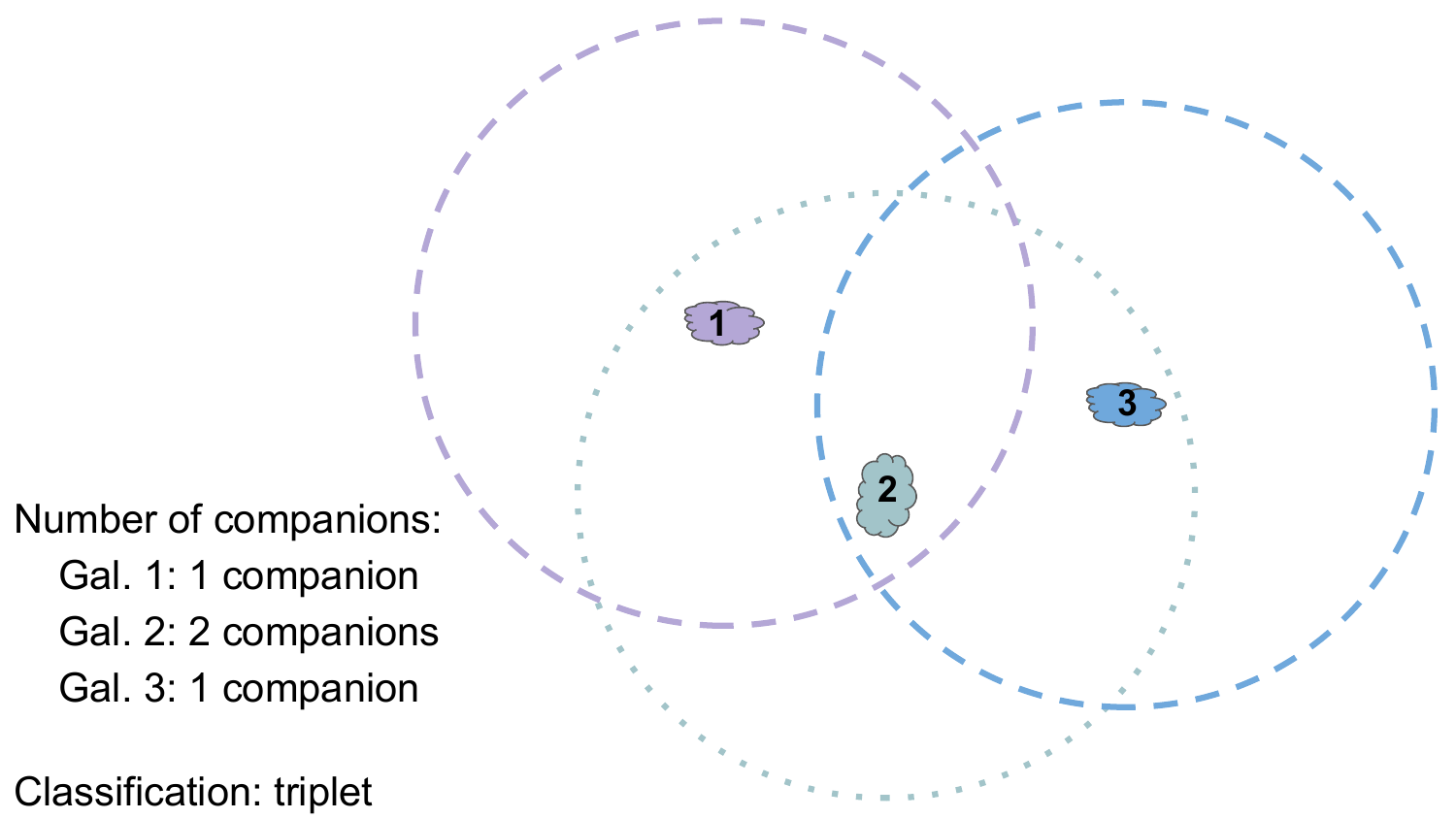}\\
        \caption{Sketch of the classification of multiples. For simplicity, we show a case where all galaxies have the same stellar mass. The three galaxies are denoted with numbers one to three, and the selection criteria (both spatial and spectral) for companions are visualized as dashed (for galaxies 1 and 3) and dotted circles (for galaxy 2), centered at each galaxy. In this case, the number of companions is one for both galaxy 1 and 3, and two for galaxy 2. We classify this system as a triplet.}
        \label{p1:fig:counting_multiples}
    \end{figure}

    To study the frequency of dwarf encounters, we aim to find dwarfs that are in close proximity to each other, i.e. that have one or more companions. We will apply the following procedure to all realizations of our drawn dwarf galaxy sample (see below). A dwarf is assigned a companion if their projected distance (\rp) and systemic velocity difference is less than a chosen threshold. We assign companions to each galaxy individually, meaning that a lower mass dwarf can be assigned a companion of a higher mass (stellar and/or \hi).
    Companions are assigned by a two-step approach, as described in the following. 
    
    In the first step, we use constant thresholds in \rp\ and \deltavsys\ for all galaxies in the sample. As we aim to compare the abundance of close-by dwarfs found through our \hi\ versus optical selection, we adopt the same criteria for the selection of pairs as in B18, corresponding to 150 kpc and 150 \kms\ in \rp\ and \deltavsys, respectively. These thresholds are chosen by B18 to correspond to the virial radius and the escape velocity at that radius of the most massive galaxies in their mock catalog ($M_{200} \sim 4 \times 10^{11}$ \msun) based on the hydrodynamical cosmological Illustris-1 simulation \citepads{2014Natur.509..177V,2015A&C....13...12N}. Hence, these thresholds serve as an upper limit to the true physical thresholds for lower mass galaxies in the sample. 

    We apply these thresholds to each dwarf, treating all others in the sample as potential companions. The number of assigned companions varies for each dwarf and may differ from the count assigned to its companions, as illustrated in Fig. \ref{p1:fig:counting_multiples}. An exception is made for dwarfs close to the spectral edges of Apertif cubes, that is, with \vsys\ < 424 + 150 = 574 \kms\ or \vsys\ > 10,170 - 150 = 10,020 \kms; these are considered as companions to other dwarfs but are not assigned their own companions to avoid underestimating the number of companions due to cube limits. This constraint is not applied to spatial edges of Apertif fields, as fields within our spatial coverage are not necessarily contiguous, thereby leaving us with a complicated sky coverage. In addition, the sensitivity within our spatial footprint is generally highly non-uniform even within individual fields (see Sect. \ref{p1:sec:apertif_data}). A detailed assessment of these sensitivity variations is beyond the scope of this paper and will be addressed in a future study. For this work, not addressing these limitations can potentially only lead to a small suppression in the number of paired galaxies.

    In the second step, we aim to make a more physical selection by scaling the thresholds in \rp\ and \deltavsys\ with respect to the halo masses of the paired dwarfs found in the first step. 
    We choose the new thresholds based on the dark matter halo masses obtained from the SHMR from the DarkLight model \citepads{2024arXiv240815214K}, and by applying the same physical definition as in the first step (virial radius and the escape velocity at that radius)\footnote{This corresponds to 144 kpc and 153 \kms\ for stellar masses of $5\times 10^9$ \msun\ when taking the SHMR from DarkLight. The 3 \kms\ increase in the spectral threshold when compared to the initial constant thresholds is smaller than half the channel width of Apertif, and would hence not make a significant difference in our result.}.
    We apply these thresholds by going through the list of galaxies who had one or more companions assigned in the first step, and revisit the assignment of each companion again using the new scaled threshold. In each comparison between the reference galaxy and its initially assigned companion, the new scaled thresholds are calculated using the stellar mass of the more massive galaxy of the two.

    Both these procedures are implemented on a series of 500 stochastic realizations of our dwarf galaxy sample, each derived by randomly drawing stellar masses assuming Gaussian error-bars in logarithmic space, and then filtering out galaxies with \mstar\ $< 5\times 10^9$ \msun. Hence, for consistency, the halo mass used for the assignment of scaled thresholds is based on the drawn stellar mass of the galaxy in each realization. All the statistical quantities presented, along with their uncertainties, are determined using the mean values and the standard deviations between all 500 realizations. We will refer to these realizations as the drawn dwarf galaxy sample (see Fig. \ref{p1:fig:flowchart}), keeping in mind that the stellar mass of any one galaxy can change between realizations.

    We report the number of pairs, triplets and quadruplets found in our drawn dwarf galaxy sample in Table \ref{p1:tab:number_of_multiples}. Multiples (pairs, triplets and quadruplets) were classified as sketched in Fig. \ref{p1:fig:counting_multiples}, where galaxies were taken to be part of the same multiple if they are sharing a common companion. Taking the mass-scaled (constant) thresholds, fraction of galaxies in a multiple of $n$ galaxies ($f_n = n\, N_n / N_{\rm tot}$) is around 10\% (12\%) for pairs ($n=2$, $N_n = N_{\rm pair}$), 2\% (4\%) for triplets ($n=3$, $N_n = N_{\rm trip}$), and 0.1\% (0.3\%) for quadruplets ($n=4$, $N_n = N_{\rm quad}$).

    \begin{table}
    \centering
    \caption{Total number of multiples in the drawn dwarf galaxy sample}
    \resizebox{0.5 \textwidth}{!}{
    \begin{tabular}{llllll}
    \hline \hline
    Sample & Thresh. & $N_{\rm tot}$ & $N_{\rm pair}$ & $N_{\rm trip}$ & $N_{\rm quad}$ \\
    \hline
    \multirow{2}{*}{Total} & const.& $2213 \pm 10$ & $132 \pm 3$ & $27 \pm 2$ & $1.6 \pm 0.7$ \\
                             & scaled & $ 2230 \pm 10$  & $110 \pm 4$ & $12 \pm 2$ & $0.7 \pm 0.5$ \\
    \multirow{2}{*}{Isolated} & const.& $638 \pm 13$  & $34 \pm 1$ & $6 \pm 1$ & $0.1 \pm 0.3$ \\
                               & scaled & $ 642\pm 11$ & $28 \pm 1$ & $5 \pm 1$ & $0$ \\
    \hline
    \label{p1:tab:number_of_multiples}
    \end{tabular} }
    \tablefoot{$N_{\rm tot}$ denotes the number of dwarfs for which the companions were searched for. Thresh. denotes which thresholds for the selection of companions were used. $N_{\rm pair}$, $N_{\rm trip}$, and $N_{\rm quad}$ are the number of pairs, triplets and quadruplets, respectively. Values and their errors correspond to the mean and the standard deviation between 500 realizations of the sample. Multiples are classified as illustrated in Fig. \ref{p1:fig:counting_multiples}.}
    \end{table}

\subsection{Isolation criteria}
\label{p1:sec:isolation_criteria}

    \begin{figure*}
        \centering
        \includegraphics[width = 1 \textwidth]{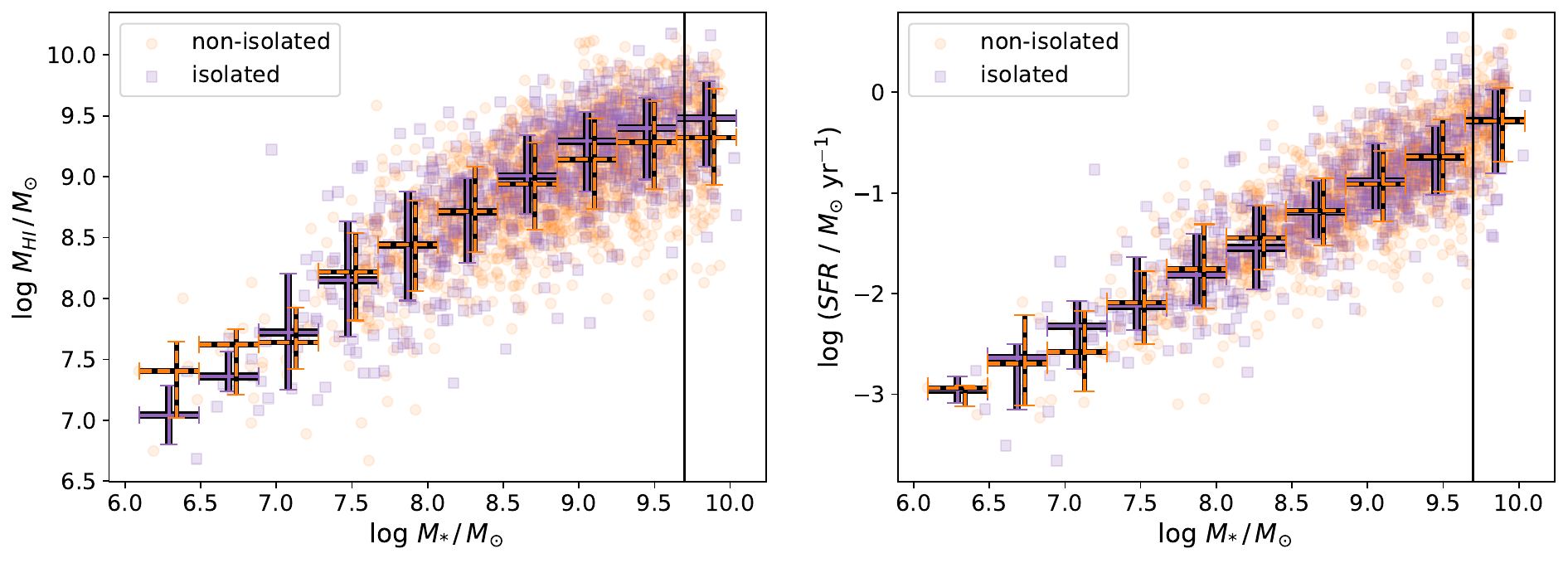}
        \caption{Comparison of galaxy properties between the isolated and non-isolated subsamples of our parent dwarf sample. \textit{Left:} \mstar\ - \mhi\ relation. Isolated galaxies are plotted as purple squares, and non-isolated as orange circles. The error bars denote the median values and percentile errors, with isolated sample denoted as full purple lines, and non-isolated as orange dashed lines. The black vertical line denotes the stellar mass limit of $5\times 10^9$ \msun. \textit{Right:} The \mstar\ - SFR relation. The markings are the same as in the left panel.}
        \label{p1:fig:mstar_mhi_SFR_isolated_vs_not}
    \end{figure*}

    To ensure a consistent comparison with the previous optical study by B18, as well as to test the impact of the Perseus-Pisces filament on our results (see Sect. \ref{p1:sec:dwarf_sample}), we adopt the isolation criteria from B18
    which ensure the dwarf galaxy is isolated from any massive galaxy (\mstar\ > $5\times 10^9$ \msun). The dwarf is not considered isolated if there is a massive galaxy that satisfies both of the following criteria: 

    \begin{enumerate}[label=(\alph*)]
        \item \deltavsys\ < 1000 \kms, where \deltavsys\ is the absolute difference in systemic velocities between the massive galaxy and the dwarf;
        \item{tidal index $\Theta$ > - 11.5, defined in \citetads{2013AJ....145..101K} as: \begin{equation}
            \Theta = \log \left( \frac{\text{\mstarmassive} [10^{11} \text{\msun}]}{r_{\rm p}^3[\text{Mpc}^3]} \right) - 10.97
        \end{equation}}
        where \mstarmassive\ is the stellar mass of the massive galaxy, and \rp\ is the projected separation between the massive galaxy and the dwarf.
    \end{enumerate}
    Criteria a) is based on the findings of \citetads{2000ApJ...536..153P} who showed that the \deltavsys\ distribution of galaxies in The Second Southern Sky Redshift Survey \citepads{1998AJ....116....1D} becomes indistinguishable from a random distribution at \deltavsys\ > 1000 \kms. On the other hand, criteria b) is based on findings from \citetads{2012ApJ...757...85G} which showed that the fraction of quenched dwarf galaxies drops to zero for projected distances larger than 1.5 Mpc from a luminous galaxy with $K$-band magnitude below $M_{K}$ < -23 (corresponding to \mstarmassive\ = 1 $\times\ 10^{11}$ \msun). This further corresponds to the tidal index of -11.5.
    Hence, these arguably stringent isolation criteria ensure dwarfs being well beyond the virial radius of any massive galaxy.

    We apply the above criteria for each dwarf from the parent dwarf galaxy sample which includes galaxies whose \mstar\ measurement is compatible at 1$\sigma$ level with our higher mass limit of $5 \times 10^9$ \msun\ (see Sect. \ref{p1:sec:dwarf_sample}). We first apply the above criteria taking detections of massive galaxies (ones outside the parent dwarf sample) from the Apertif source list. This left us with 1244 candidate isolated dwarfs. We further proceed to query NED within \rp\ < 2 Mpc and |$\Delta$ \vsys| < 1000 \kms\ around each of the remaining dwarfs, in order to find massive galaxies that might have been missed in Apertif (e.g. ones with low gas fractions) or that fall outside the Apertif footprint. 
    During this process we estimated stellar masses of NED galaxies using the W1 22.0\arcsec\ radius aperture magnitude, and applying equation 2 from \citetads{2013MNRAS.433.2946W}. This left us with 735 isolated dwarf galaxy candidates (i.e. the parent isolated sample in Fig. \ref{p1:fig:flowchart}). We note that by using a fixed aperture magnitude, we are potentially underestimating stellar masses of high mass galaxies from NED (22\arcsec\ corresponds to 7.5 kpc at our median distance of 70 Mpc). Taking the Milky Way (\mstar\ = $5 \times 10^{10}$ \msun; \citeads[e.g.][]{2016ApJ...831...71L,2016ApJ...816...42M}) as an example, approximately 80\% of its mass\footnote{Assuming exponential stellar disk with the scale length of 2.5 kpc.} is enclosed within 7.5 kpc. This underestimates the required projected separation for isolation at 1.1 Mpc, instead of 1.2 Mpc. This small difference could result in the misclassification of a galaxy as isolated. 
    It is, however, worth noting that most of the observed volume ($V$) is, and hence most of the massive galaxies are, beyond the median distance ($V \propto r^3$) where the effect is smaller. Furthermore, stellar distributions of gas-poor early-type galaxies are likely more concentrated than an exponential disk, making the effect milder than presented in the Milky Way example.
    Taking this into consideration together with the extreme strictness of the isolation criterion, we believe our approximate calculation of the stellar mass of high mass galaxies is sufficient to create a well isolated dwarf galaxy sample. Finally, within each realization of the drawn dwarf galaxy sample, we further apply the isolation criteria for every dwarf with drawn \mstar\ $< 5\times 10^9$ \msun, with regards to the ones whose drawn stellar mass exceeds this stellar mass limit. The final number of isolated dwarfs per realization is around $\sim$649 (i.e. drawn isolated sample in Fig. \ref{p1:fig:flowchart}).

    In Fig. \ref{p1:fig:mstar_mhi_SFR_isolated_vs_not}, we show the \mstar\ - \mhi\ relation and the \mstar\ - SFR relation of our isolated vs. non-isolated galaxies in the parent dwarf sample. Generally, both relations show very good consistency between the two samples. The only systematic offset between the two populations is seen at the highest masses in the \mstar\ - \mhi\ relation, with isolated galaxies being systematically more gas-rich. This is expected as higher-mass dwarfs can retain their gas content in closer proximity to high-mass galaxies, thereby still being easily detectable in \hi.
    However, we see no offset in the SFRs between the two populations, implying that the impact of the environment only influenced the non-starforming gas reservoir (e.g. at the galaxies' outskirts, by gas stripping). 
    The apparent larger \hi\ masses for non-isolated dwarfs at very low masses is not statistically significant due to the low number of objects ($\leq$ 8 per bin in either sample). Hence, while there are signatures of environmental impacts on the highest mass galaxies in our parent dwarf sample, it is unlikely that we have strong contamination from dwarfs in direct close-by interactions with a high-mass galaxy, meaning that our total sample is still mostly dominated by reasonably well isolated dwarfs.

    Interestingly, the number of multiples found in the isolated drawn galaxy sample does not deviate significantly from the number of multiples in our total drawn sample, as seen from Table \ref{p1:tab:number_of_multiples}. The absolute fractional difference in the number of galaxies in multiples ($|f_n^{\rm total} - f_n^{\rm isolated}|$) for mass-scaled (constant) thresholds are 1.1\% (1.3\%) for pairs, 0.7\% (0.8\%) for triplets, and 0.1\% (0.2\%) for quadruplets.

\section{Results}
\label{p1:sec:results}
\subsection{Frequency of dwarf galaxy multiples}
\label{p1:sec:multiples_Apertif}

    \begin{figure}
        \centering
        \includegraphics[width = 0.5 \textwidth]{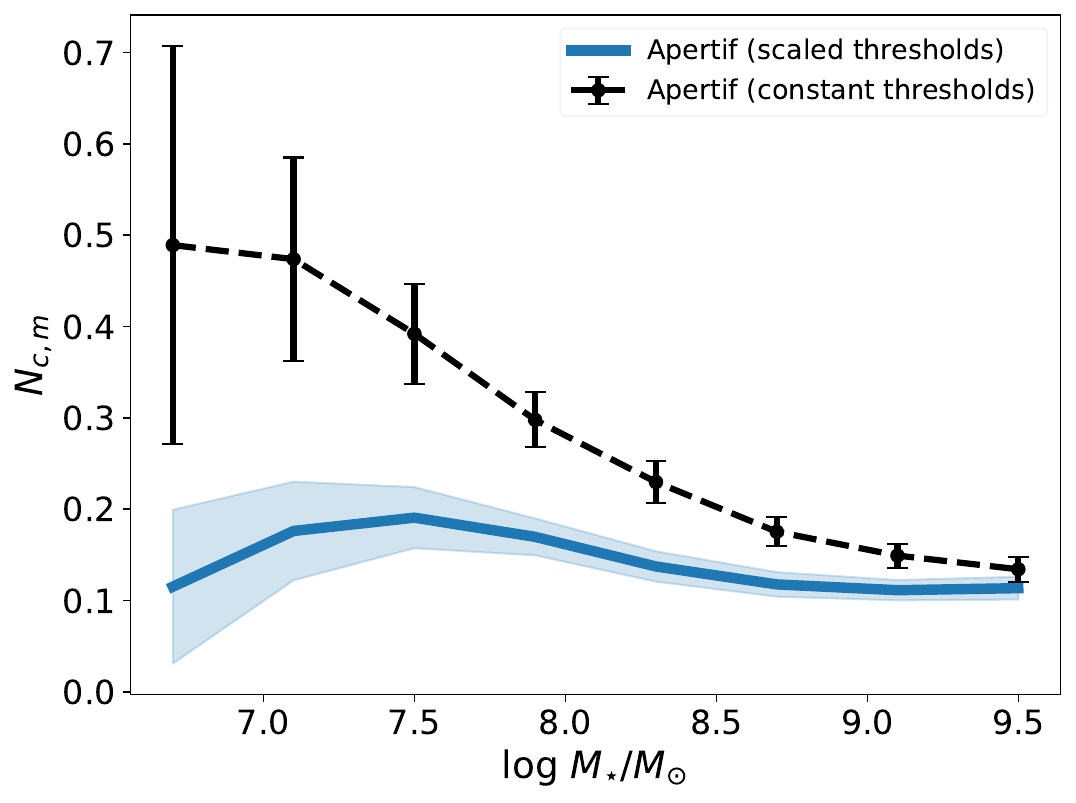}
        \caption{Mean number of companions (of any stellar mass) per dwarf galaxy in the logarithmic stellar mass bin, normalized by the number of dwarfs in the same mass bin. Values represent the mean between different realizations of the sample in each mass bin. The black line represents results obtained using constant thresholds in projected distance and systemic velocity difference (150 kpc and 150 \kms, respectively) for selection of companions, while the blue line represent results obtained using the mass-scaled thresholds (see Sect. \ref{p1:sec:selection_companions}). Both lines represent results for the total drawn dwarf galaxy sample. Shaded regions and error bars represent 1$\sigma$ errors.}
        \label{p1:fig:N_cm_constant_scaled}
    \end{figure}

    \begin{figure*}
        \centering
        \includegraphics[width = 1 \textwidth]{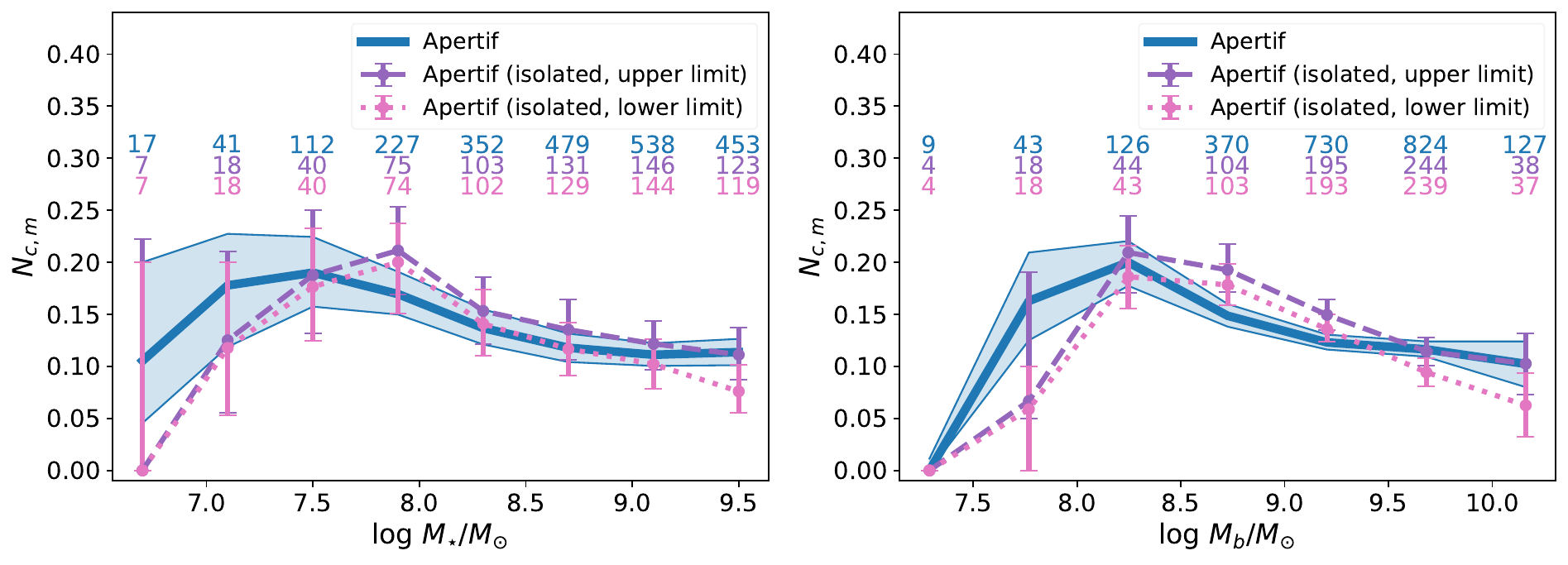}
        \caption{
        Mean number of companions (of any stellar/baryonic mass) per dwarf in logarithmic stellar (left) or baryonic (right) mass bins, normalized by the number of dwarfs per bin. Values represent the median between different realizations of the sample in each mass bin. All lines are obtained using mass-scaled thresholds (see Sect. \ref{p1:sec:selection_companions}): the blue line represents the total drawn dwarf galaxy sample, the purple dashed line the isolated drawn sample including split multiples (upper limit), and the pink dotted line the isolated drawn sample excluding split multiples (lower limit). Shaded regions and error bars indicate 1$\sigma$ uncertainties. The number of galaxies in each bin is shown at the top, with font colors and order matching the corresponding lines in the legend.
        }
        \label{p1:fig:N_cm_isolated}
    \end{figure*}

    In this section, we investigate how the average number of dwarf galaxy companions depends on the stellar mass of the dwarf, or in other words, how many companions a dwarf of a specific stellar mass is expected to have on average.
    We show our results in terms of the mean number of companions per galaxy per stellar mass bin ($N_{c,m}$; \citeads{2000ApJ...536..153P}, B18), calculated by summing the number of companions (of any stellar mass) of all dwarfs corresponding to the stellar mass bin, and by dividing by the total number of dwarfs in that mass bin.  

    In Fig. \ref{p1:fig:N_cm_constant_scaled}, we compare our results obtained by applying constant thresholds for projected distance and systemic velocity difference for the selection of companions (from the first step in our analysis, see Sect. \ref{p1:sec:selection_companions}), and the ones obtained after applying the mass-scaled thresholds on our total drawn dwarf galaxy sample (without applied isolation). 
    In both cases the mean number of companions rises as we go down toward lower stellar masses, with the rise being significantly milder with the use of the mass-scaled thresholds. This difference corresponds to potential non-physical associations present when using constant thresholds (scaled to the highest stellar masses in the sample) for companion selection of lower mass dwarfs. This trend is also found in simulations, for example, \citetads{2022MNRAS.509.5918H} and \citetads{2024ApJ...962..162C}. Hence, using mass-scaled thresholds is necessary to systematically constrain the fractions of dwarf companions in Apertif, and we will use them in the rest of the paper, unless stated otherwise.

    In the mass-scaled case, the slight increase in the mean number of companions toward lower stellar masses is still expected due to our stellar mass cut on the sample. At the highest masses in our sample, we are, for example, only counting half of the major pairs these galaxies are a part of, the ones in which they are preferentially the hosts (most massive in the pair / multiple). Similarly, at the lowest masses, we are only counting pairs in which the reference galaxy is a satellite (not the most massive in the multiple), contributing to the slight turndown in the number of companions. Finally, for dwarfs in the middle of our mass range, we are counting both cases, when they are either the host or the satellite. Despite these expectation, this trend is mild in our sample, as the mean number of companions with scaled thresholds is consistent with a constant value within 2$\, \sigma$. The exact shape of the trend, however, inherently depends on the slope of the adopted SHMR (Eq. \ref{p1:eq:SHMR}) which is used to define the mass-scaled thresholds. Shallower SHMRs would give a slightly stronger increase in the mean number of companions toward lower stellar masses. In summary, by adopting the DarkLight SHMR, we find that the mean number of companions per dwarf (taking the whole mass range) is (12.8 $\pm$ 0.4)\% in our sample.

    The comparison of results for our total drawn sample and our isolated drawn sample are shown in Fig. \ref{p1:fig:N_cm_isolated}, where we show binnings both in stellar and baryonic masses. 
    In this plot we have used the median and the percentile errors in each stellar (left panel) and baryonic (right panel) mass bin due to low number statistics in the lowest mass bins. We plot two cases for the isolated sample as some multiples ($\sim$ 10, depending on the realization) would end up being split up after applying the criteria (i.e. one galaxy passed the isolation criteria while the second has not). Given that the multiples are chosen to be close-by and our isolation criteria are strict, we prefer to either include or exclude the whole multiple when counting. Hence, we show a case with all galaxies within a split multiple corresponding to our upper limit (purple) and a case with none of the galaxies in a split multiple corresponding to our lower limit (pink). The mean number of companions in both cases are fully consistent with the result from the total drawn sample, but seem to show a steeper increase toward lower stellar (and baryonic) masses at higher mass bins (even though still consistent with a constant distribution at 2$\sigma$). This might indicate an influence of the large scale structure on the groupings of dwarf galaxies. However, the general consistency of the isolated and full sample shows us that the impact (if real) on our sample is minor, which is consistent with our \hi-based selection that is already biased to isolated galaxies.

\subsection{Star formation rate of paired galaxies}
\label{p1:sec:sfr_pairs}

\begin{figure*}
        \centering
        \includegraphics[width = 1 \textwidth]{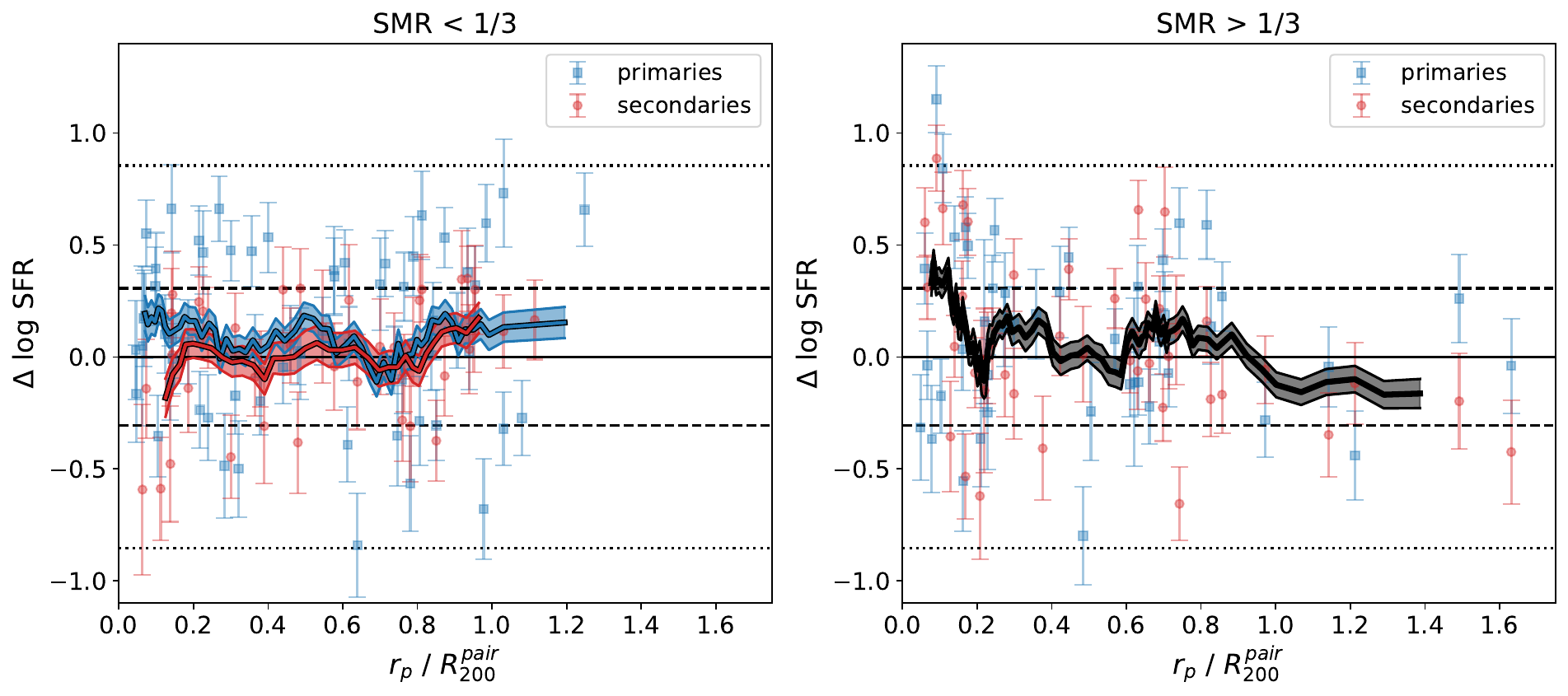}
        \caption{Difference in the SFRs of paired galaxies with respect to the unpaired control sample versus the pair separation scaled by the virial radius of the pair. \textit{Left:} Pairs with stellar mass ratios (SMRs) below 1/3. Primaries correspond to the more massive galaxy in a pair and are denoted as blue squares, while secondaries (less massive in a pair) are denoted as red circles. Blue and red lines correspond to the running means for primaries and secondaries, respectively, with 1$\sigma$ errors on the means denoted as shaded regions. Black dashed and dotted horizontal lines denote the scatter of 1$\sigma$ and 2$\sigma$ of unpaired galaxies. \textit{Right:} Pairs with SMRs above 1/3. The notation is the same as in the left panel, except that the running mean (black line) is calculated by taking primary and secondary galaxies together.
        }
        \label{p1:fig:multiples_properties_dMS_rp}
    \end{figure*}

    As shown in previous studies \citepads[e.g.][]{2024ApJ...963...37K,2025ApJ...980..157H}, close galaxy encounters can enhance and suppress star formation in dwarf galaxies. Different mechanisms have been proposed for the SFR enhancement in (gas-rich) high- and low-mass galaxies 
    \citepads{2025ApJ...980..157H}. To explore if such trends are present in our sample, we select pairs of dwarfs identified in at least one realization of our total drawn galaxy sample using constant thresholds (from the first step of our analysis, see Sect. \ref{p1:sec:selection_companions}). We use these looser thresholds to have better statistics, noting that they are relatively strict compared to previously used ones in similar works (e.g. 250 kpc and 500 \kms\ used in \citeads{2025ApJ...980..157H}). The total number of pairs obtained is 147.
    We consider only galaxies in pairs and not higher multiples for easier interpretation. In this section, we use the original stellar mass measurement (corresponding to the parent galaxy sample) instead of the drawn value.
    
    For each galaxy in a pair, we assign a control sample of five or more unpaired dwarfs (ones with no companion in any realization) that have both stellar and \hi\ masses within 0.1 dex of the paired dwarf's. In 47 cases for which there were less than five single dwarfs satisfying the criteria, we excluded the paired dwarf from the analysis. Following the approach of \citetads{2025ApJ...980..157H}, we define the SFR offset (\dSFR) as the difference between the SFR of the paired dwarf and the median of the SFRs of its control sample. The error on this median is obtained by bootstrapping SFRs of the control sample. Furthermore, to estimate the significance of any possible offsets, we randomly select around 250 unpaired galaxies and repeat the above procedure where we assign them their own control sample and calculate their \dSFR. We then use 1$\sigma$ and 2$\sigma$ scatter of the unpaired dwarfs for our comparison to paired dwarfs. 

    Fig. \ref{p1:fig:multiples_properties_dMS_rp} shows the \dSFR\ with respect to the projected distance of individual galaxies in pairs, scaled by the virial radius of the pair. As in \citetads{2025ApJ...980..157H}, this virial radius is calculated using the halo mass of the pair, which we estimate by summing the halo masses (see Sect. \ref{p1:sec:bary_halo_masses}) of individual galaxies. 
    We distinguish the primary and secondary galaxies as being the more and less massive (in stellar mass) in the pair, respectively, and separate our sample to high and low stellar mass ratio (SMR = \mstar$^{\rm secondary}$ / \mstar$^{\rm primary}$) pairs based on the threshold of 1/3. In the low (SMR<1/3) SMR case, we calculate running means for primary and secondary galaxies separately, while for the high (SMR>1/3) SMR case we do not make this distinction as galaxies within a pair are comparable in mass. 
    We calculate the running means in the SMR > 1/3 case using the kernel size of 10, corresponding to the mean span in \rp/\Rvirpair\ (in the running mean calculation) of 0.07 for scaled separations below 0.3, 0.19 between 0.3 - 0.6, and 0.13 between 0.6 - 0.9. To roughly match the same mean \rp/\Rvirpair\ span and similar statistical significance in the SMR < 1/3 case, we use the kernel size of 8, corresponding to the mean span of 0.16 and 0.20 for primaries and secondaries, respectfully.
    Errors on the running mean are obtained by bootstrapping the \dSFR\ offsets of paired galaxies and taking the mean and the standard deviation of the running mean between $10^4$ realizations.

    In both high and low SMR cases, the scatter of paired galaxies (both primaries and secondaries) is comparable to the scatter of unpaired dwarfs at all scaled separations. However, for high SMRs and at scaled separations below 0.2, there are a few paired dwarfs showing SFR enhancements above the $2\sigma$ scatter of unpaired galaxies, up to around 1 dex enhancement compared to their unpaired control samples. 
    There is also a systematic increase of the running mean at these separations, reaching 0.33 dex above the mean of the unpaired dwarfs (zero enhancement).
    At larger separations, the density of the points decreases, so the observed trend is less well constrained. It is, however, clear that we do not see any significant SFR suppression at any scaled separation in this case.

    For low SMRs, we see no significant trend between primary and secondary galaxies with respect to the scaled separation, but there is a hint of the SFR enhancement of primaries and SFR suppression of secondaries at the lowest separations (below 0.2). However, primaries do seem to show a systematic SFR enhancement at almost all separations ($\sim 0.09$ dex, on average).

\section{Discussion}
\label{p1:sec:discussion}

\subsection{Frequency of pairs in optical vs. our \hi\ sample}
\label{p1:sec:comparison_w_optical}

        \begin{figure*}
        \centering
        \includegraphics[width = 0.95 \textwidth]{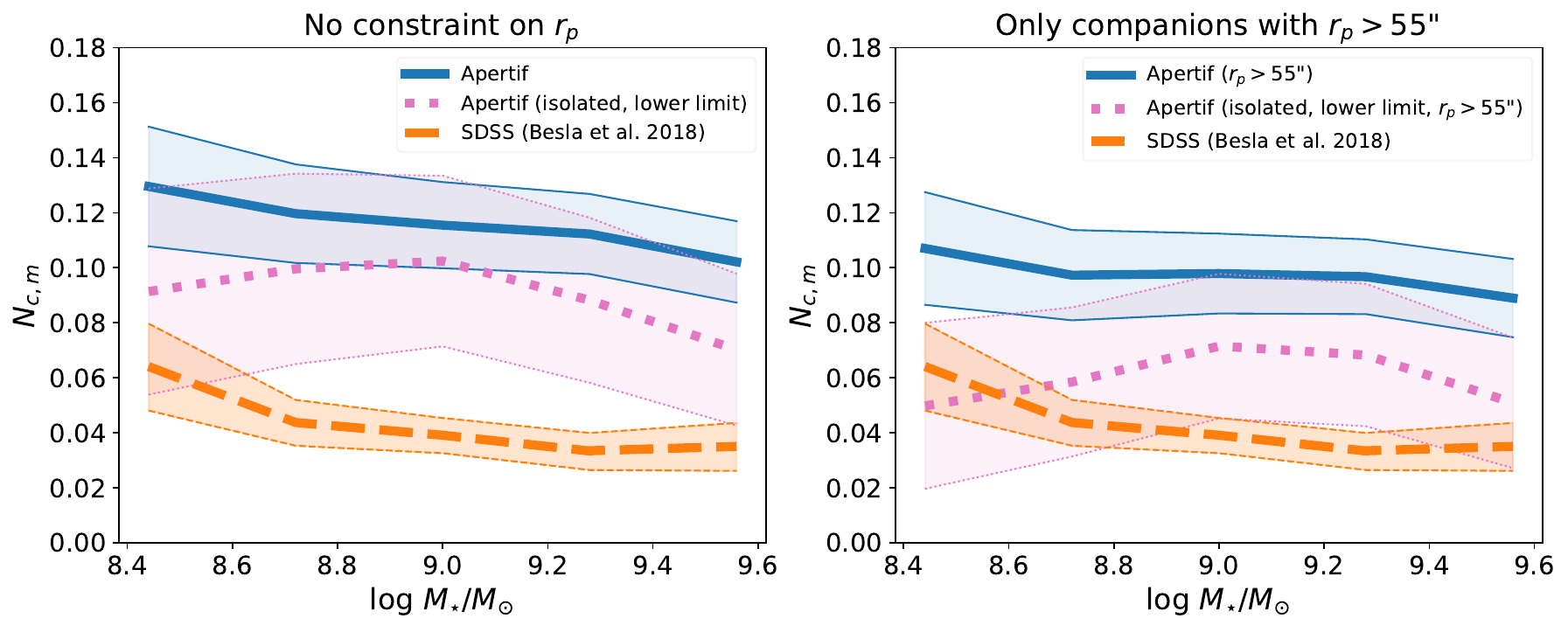}\\
        \caption{Comparison of the mean number of companions per dwarf between the Apertif and the SDSS samples. \textit{Left:} The case with no constraint on the projected radius between companions in Apertif. Blue line represents the result for the total drawn galaxy sample, pink dotted line results from the isolated drawn sample by taking the lower limit (see Sect. \ref{p1:sec:multiples_Apertif}), and orange dashed line represents results from B18. Shaded regions represent 1$\sigma$ errors. \textit{Right:} The same as in the left panel, but excluding all companions in Apertif for which \rp\ < 55\arcsec.}
        \label{p1:fig:N_cm_Besla}
    \end{figure*}

    We compare the mean number of companions in our sample to results from B18 who studied the number of dwarf multiples found in SDSS spectroscopic data within a stellar mass range of \smrBesla. They have used constant thresholds in their selection of companions, exactly corresponding to the first step in our selection (Sect. \ref{p1:sec:selection_companions}), and have applied the same isolation criteria that we employ in our sample (Sect. \ref{p1:sec:isolation_criteria}). They only have one additional criterion that they impose in their selection of companions, namely that the projected radius between pairs is \rp\ > 55\arcsec\ in order to avoid incompleteness due to fiber collisions. In the following comparison, we cut our drawn dwarf galaxy samples to the same stellar mass range to ensure a consistent comparison and explore cases with and without applying the additional criterion on \rp.
    
    The comparison between B18 and our results can be seen in Fig. \ref{p1:fig:N_cm_Besla}, where in the left panel we do not apply the additional criterion on the projected radius (\rp\ > 55\arcsec), and in the right panel we do. 
    For each of these cases, we show results with and without applying the isolation criteria to our sample (Sect. \ref{p1:sec:isolation_criteria}). We include the latter case in the comparison with B18 as our full sample contains a comparative number of galaxies in the sample as in B18 ($\sim$1660 in our Apertif sample, and $\sim$1910 in SDSS) and is likely still dominated by field galaxies (see Sect. \ref{p1:sec:isolation_criteria}), thereby making for a more statistically consistent comparison. For the case with applied isolation, we take the lower limit where we exclude pairs even if only one dwarf did not pass the isolation criteria (see Sect. \ref{p1:sec:multiples_Apertif}). 
    
    Overall, in all these comparisons our Apertif (\hi\ selected) sample returns a higher mean number of companions per dwarf galaxy. In the case without applying the additional selection criteria (left panel), the full Apertif sample is remarkably higher in all mass bins. This remains true even when the isolation criteria is applied to the sample, although the reduced number of galaxies in the Apertif sample ($\sim$443 galaxies) means it becomes consistent within 2$\sigma$. 
    When considering the additional criteria to avoid close companions (\rp\ > 55 \arcsec; right panel) and applying the same isolation criteria, the two samples become formally consistent, although with our values still systematically higher across most of the mass range. However, this is limited by the small number of galaxies remaining in the Apertif sample leading to large uncertainties. In comparison, when taking our full sample, the mean number of companions is ($11.6\pm 0.5$)\%, consistently higher than in B18, and also remains higher than their predicted value of 6\% for a stellar mass complete survey.

    A possible explanation for the offset between our \hi\ selected sample and the B18 sample could be the additional galaxy population we are probing. Optical spectroscopic studies are biased against detecting low surface brightness galaxies, while in \hi\ we are able to easily detect all gas-rich dwarfs independently of their stellar content and obtain the redshift information at the same time. Hence, cases in which the interaction disturbs the morphology of an interacting dwarf (potentially lowering its surface brightness) might be missed in optical surveys, thereby leading to a lower fraction of multiples.
    
    Our result suggests that \hi\ surveys are able to find three times higher percentage of multiples than optical surveys. Further studies with larger samples are needed to robustly constrain how different selections impact the retrieved number of multiples and how significant the offset we are finding truly is. It is, however, clear that \hi\ surveys allow identification of more close pairs by not suffering from fiber collisions and thus provide a crucial window for quantifying the actual number of dwarf galaxy multiples in the field as well as studying the closest part of their interactions. These results can be directly compared to the clustering of galaxies in the low mass regime of current state-of-the-art simulations, thereby constraining the structure formation in current models, the implementation of sub-grid physics (especially during dwarf-dwarf interactions), and, hence, galaxy evolution at small scales. Finally, as \hi\ surveys are naturally unbiased by the stellar content, the impact that the interaction has on the dwarf (e.g. lowering the surface brightness due to tidal effects or enhancing the surface brightness by enhancing the SFR), will not impact the selection of the dwarf, making the estimate of pair fractions more robust than in the optical.

\subsection{Properties of \hi\ selected dwarf pairs}
\label{p1:sec:comparison_w_wallaby}

    A recent work by \citetads{2025ApJ...980..157H} studied how close encounters of gas-rich galaxies influence their SFRs. They selected galaxy pairs from the WALLABY \hi\ survey, using thresholds in projected radius and velocity difference of 250 kpc and 500 \kms, respectively. They also excluded higher order multiples by imposing that neither galaxy in a pair has another companion within 250 kpc and 500 \kms, and have made sure none of the pairs are inside known galaxy groups to ensure isolation. Their sample spans the stellar mass range of $10^{7.6}$ to $10^{11.2}$\msun. For each paired galaxy, they have made a control sample by matching in stellar and \hi\ masses, and redshift, from a control pool of galaxies that have no imposed constraints (including no isolation constraint).

    They separate their sample to low- and high-mass galaxies based on the stellar mass limit of $10^9$\msun, but allow their low-mass galaxies to be paired with high-mass ones.
    For their low-mass galaxy sample, the SFRs initially decrease below \rp $\lesssim$ 1 \Rvirpair\ with maximum SFR suppression of $\sim$0.22 dex (at 1.4$\sigma$ significance) at scaled separations of 0.6. Only at separations \rp $\lesssim 0.2$ \Rvirpair, they find a steep SFR enhancement of magnitude $\sim$0.22 dex.
    As seen in Sect. \ref{p1:sec:sfr_pairs}, the high SMR pairs in our sample show the SFR enhancement at separations \rp $\lesssim 0.2$ \Rvirpair, but no clear indication of systematic SFR suppression.
    This holds true even if we apply the upper stellar mass limit of $10^9$ \msun, as in \citetads{2025ApJ...980..157H}. However, in the low SMR pairs we do find a tentative SFR suppression for secondaries at separations below 0.2, indicating that the large difference in mass might be needed in order to suppress the star formation in the lower mass galaxy.
    We note that, differently from \citetads{2025ApJ...980..157H}, we do not include dwarf-high mass galaxy interactions in our analysis. Hence, such interactions can potentially have a much stronger influence on the star-formation activity of dwarf galaxies in their work, thereby inducing stronger and potentially traceable trends.

\subsection{Impact of merged detections}
\label{p1:sec:impact_of_merged}

    \begin{figure}
        \centering
        \includegraphics[width = 0.48 \textwidth]{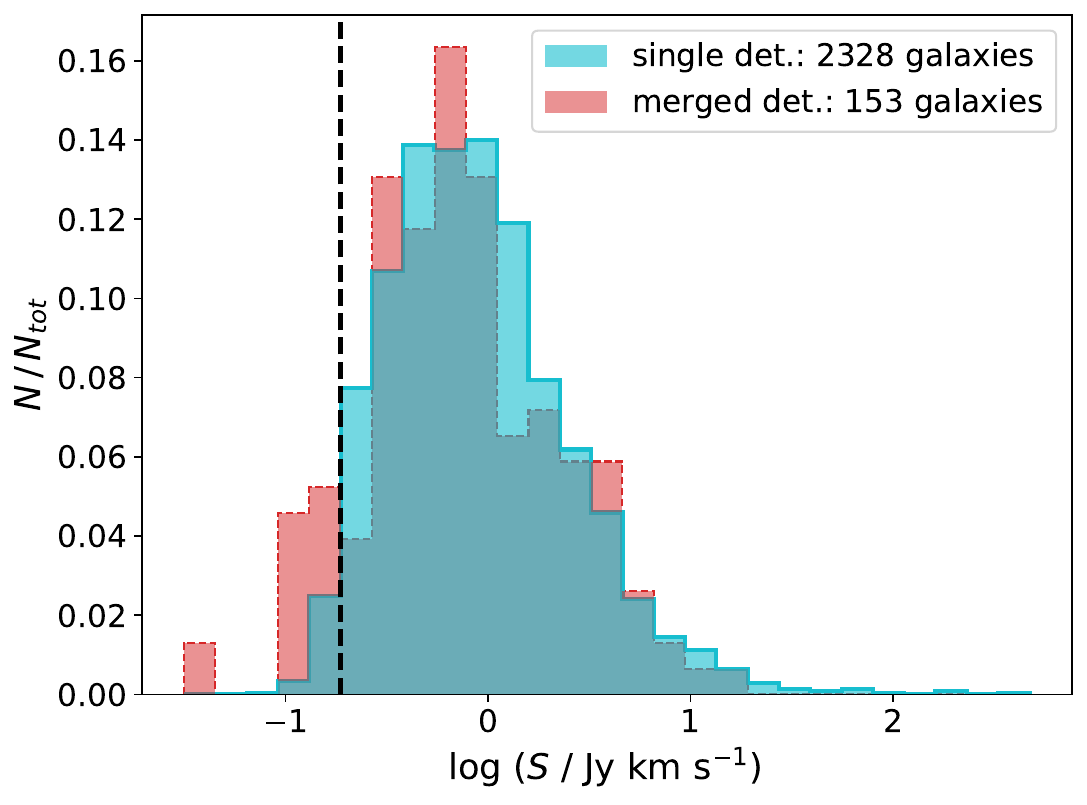}\\
        \caption{Flux distributions for galaxies found in merged detections (red) versus separate detections (cyan) in our parent dwarf galaxy sample. Histograms are normalized to the number of galaxies in each sample (reported in the legend). Black dashed vertical line denotes the flux limit of 0.187 Jy \kms\ below which there is an excess of galaxies detected in merged detections.}
        \label{p1:fig:merged_vs_non_merged_flux}
    \end{figure}

     \begin{figure}
        \centering
        \includegraphics[width = 0.49 \textwidth]{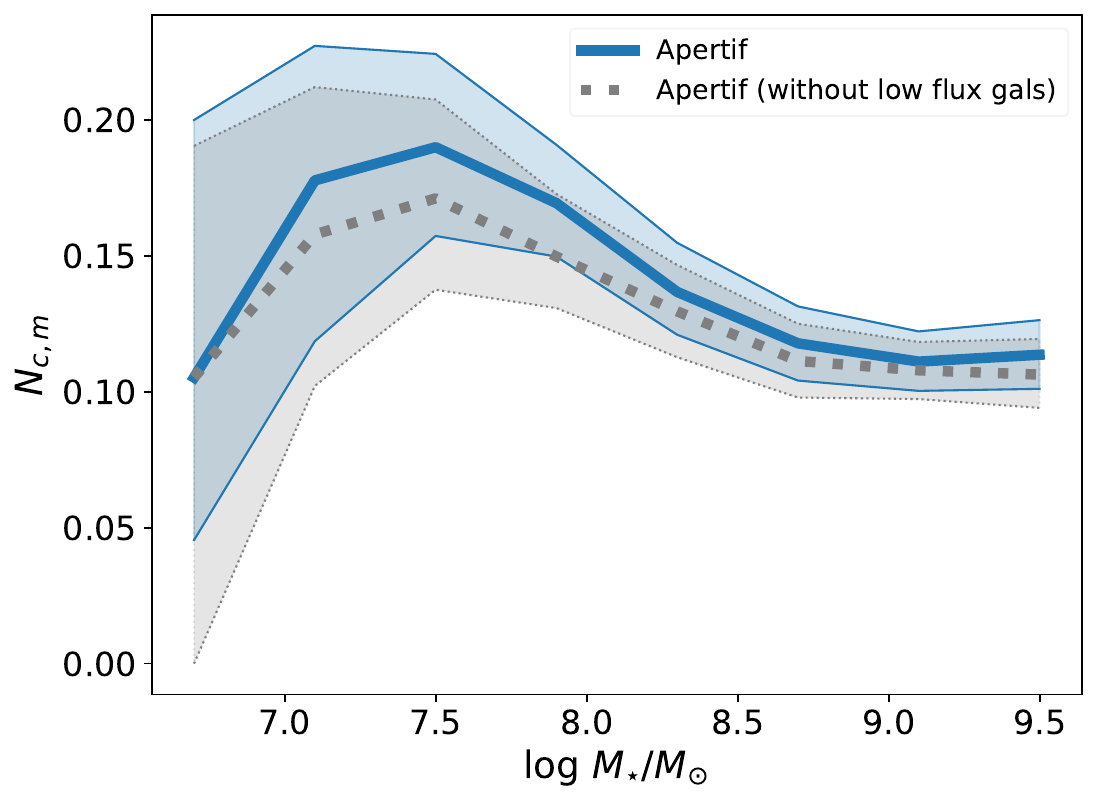}
        \caption{Mean number of companions (of any stellar mass) per dwarf in the logarithmic stellar mass bin, normalized by the number of dwarfs in the same mass bin. Values represent the median between different realizations of the sample in each mass bin. Blue line represents results for the total drawn dwarf galaxy sample and gray dotted line results obtained by excluding all sources below the 0.187 Jy \kms\ threshold in the \hi\ flux. Shaded regions represent 1$\sigma$ errors.}
        \label{p1:fig:N_cm_general}
    \end{figure}

    As described in Sect. \ref{p1:sec:producing_sample}, the parent dwarf galaxy sample contains dwarf galaxies which were detected as part of a merged \hi\ detection (155 out of 2481 galaxies in the sample). There is a consideration that some merged detections might include emission of dwarf galaxies that would have been missed as an independent \hi\ detection due to being individually too faint (i.e. close to or below the detection limit). If true, this would possibly bias our results toward a higher mean number of companions per dwarf.

    As a first estimate of the impact of potential biases merged detections could have on our sample, we plot the distribution of \hi\ fluxes of galaxies in our parent dwarf sample in Fig. \ref{p1:fig:merged_vs_non_merged_flux}, 
    distinguishing between those detected individually and those found in merged detections. If there were no biases in our sample, we would expect these two distributions to have the same shape. While the distributions are indeed similar in shape, there seems to be a small tail toward lower fluxes for the case of merged detections. This indicates that some galaxies in our sample would possibly not be detected as an individual source. 
    To estimate the impact of this on our results, we exclude all galaxies (corresponding to either single or merged detections) in the lowest five flux bins and rerun our analysis. The result is shown as gray dotted line in Fig. \ref{p1:fig:N_cm_general}. The result is completely consistent with our original run, with only 1\% offset (on average) toward lower mean number of companions.
    
    Hence, the true number of companions is most likely close to our original result. The robust quantification of any potential biases in our sample (e.g. by repeating the source finding procedure with mock sources) is, however, beyond the scope of this work.

\section{Conclusions}
\label{p1:sec:conclusions}

    In this work, we have for the first time presented the frequency of dwarf galaxy multiples found in an interferometric, untargeted \hi\ survey. Our parent dwarf galaxy sample is based on detections from the Apertif \hi\ survey and spans the stellar mass range between $\sim 10^6$ \msun\ and $\sim 5\times 10^9$ \msun\ (including higher mass galaxies within 1$\sigma$ of this threshold), containing 2481 candidate dwarf galaxies. We searched for paired galaxies by imposing thresholds in the projected radius and the difference in systemic velocity using two approaches. Firstly, we applied the constant thresholds corresponding to $r_{\rm p} < 150 $ kpc and $| \Delta V_{\rm sys} | < 150 $ \kms\ for all galaxies independently of their stellar mass, in analogy to the previous work based on optical spectroscopy by B18. In the second step, we applied mass-scaled thresholds based on the SHMR from \citetads{2024arXiv240815214K} and corresponding to the virial radius and the escape velocity at that radius. We applied a strict isolation criterion to our sample to ensure consistent comparison with previous works. Finally, we have studied the SFR enhancement in pairs found in our first step (using constant thresholds) with respect to the projected separation of the pair scaled by the virial radius of the pair. We summarize our findings in the following:
    \begin{itemize}
        \item We find that the mean number of companions per dwarf is $\sim 12.8$\% in our \hi\ selected dwarfs sample (Fig. \ref{p1:fig:N_cm_constant_scaled}). This is the first quantification of the fraction of dwarf galaxy multiples in an untargeted \hi\ survey. 
        
        \item We do not find large differences between the total \hi-detected dwarf galaxy sample and a highly isolated subsample in either global galaxy properties, or the mean number of companions per dwarf (Figs. \ref{p1:fig:mstar_mhi_SFR_isolated_vs_not} and \ref{p1:fig:N_cm_isolated}). This indicates that an \hi-selected dwarf galaxy sample is already well isolated, in line with previous studies from the literature.
        
        \item We are able to detect a three times higher fraction of dwarf galaxy pairs in the Apertif \hi\ survey than is found in the SDSS optical spectroscopic survey (Fig. \ref{p1:fig:N_cm_Besla}). 
        This illustrates the power of \hi\ for finding dwarf pairs, especially close pairs where fiber collisions are a strong limitation for existing optical spectroscopic surveys.
        
        \item Pairs of galaxies with comparable stellar masses (stellar mass ratios between 1/3 - 1) in Apertif show SFR enhancement at projected separations $\lesssim$0.2 \Rvirpair\ of the order of $\sim0.33$ dex. We also find a hint of SFR suppression of less massive galaxies in pairs with larger mass difference (stellar mass ratios below 1/3), indicating that a significant difference in mass is needed for the suppression of the SFR in dwarf galaxies.
    \end{itemize}

    This study exemplifies the power of untargeted, interferometric \hi\ surveys in robustly constraining the true frequency of dwarf galaxy multiples in observations. The frequency of galaxy encounters, together with studying the influence of such encounters on galaxy properties, is crucial for disentangling galaxy's secular evolution from their environment.
    Hence, our results could serve as a direct test of galaxy evolution models by comparison to the current state-of-the-art simulations in both frequency of dwarf multiples and properties of interacting galaxies.

\section*{Data availability}

% The parent dwarf galaxy sample used in this work is publicly available at the CDS via anonymous ftp to \url{cdsarc.u-strasbg.fr} (130.79.128.5) or via \url{http://cdsweb.u-strasbg.fr/cgi-bin/qcat?J/A+A/}.

The parent dwarf galaxy sample used in this work will be publicly available at the CDS via anonymous ftp to \url{cdsarc.u-strasbg.fr} (130.79.128.5) or via \url{http://cdsweb.u-strasbg.fr/cgi-bin/qcat?J/A+A/} upon publication.

\begin{acknowledgements}
We thank the anonymous referee for valuable comments that improved this manuscript. KMH, JG, and SSE acknowledge financial support from the grant CEX2021-001131-S funded by MCIN/AEI/ 10.13039/501100011033 from the coordination of the participation in SKA-SPAIN funded by the Ministry of Science and Innovation (MCIN); and from grant PID2021-123930OB-C21 funded by MCIN/AEI/ 10.13039/501100011033 by “ERDF A way of making Europe” and by the "European Union". Authors KMH, BŠ, JG, SSE acknowledge the Spanish Prototype of an SRC (SPSRC) service and support funded by the Ministerio de Ciencia, Innovación y Universidades (MICIU), by the Junta de Andalucía, by the European Regional Development Funds (ERDF) and by the European Union NextGenerationEU/PRTR. The SPSRC acknowledges financial support from the Agencia Estatal de Investigación (AEI) through the "Center of Excellence Severo Ochoa" award to the Instituto de Astrofísica de Andalucía (IAA-CSIC) (SEV-2017-0709) and from the grant CEX2021-001131-S funded by MICIU/AEI/ 10.13039/501100011033. KMH and JMvdH acknowledge funding from the ERC under the European Union’s Seventh Framework Programme (FP/2007–2013)/ERC Grant Agreement No. 291531 (‘HIStoryNU’). This work has received funding from the European Research Council (ERC) under the Horizon Europe research and innovation programme (Acronym: FLOWS, Grant number: 101096087).

This work makes use of data from the Apertif system installed at the Westerbork Synthesis Radio Telescope owned by ASTRON. ASTRON, the Netherlands Institute for Radio Astronomy, is an institute of the Dutch Research Council (“De Nederlandse Organisatie voor Wetenschappelijk Onderzoek”, NWO). 
This publication uses data generated via the Zooniverse.org platform, development of which is funded by generous support, including a Global Impact Award from Google, and by a grant from the Alfred P. Sloan Foundation. This work made use of Astropy:\footnote{http://www.astropy.org} a community-developed core Python package and an ecosystem of tools and resources for astronomy \citepads{astropy:2013, astropy:2018, astropy:2022}. This research has made use of the NASA/IPAC Extragalactic Database (NED),
which is operated by the Jet Propulsion Laboratory, California Institute of Technology,
under contract with the National Aeronautics and Space Administration.
This publication makes use of data products from the Wide-field Infrared Survey Explorer, which is a joint project of the University of California, Los Angeles, and the Jet Propulsion Laboratory/California Institute of Technology, funded by the National Aeronautics and Space Administration.
This work is based in part on observations made with the Spitzer Space Telescope, which was operated by the Jet Propulsion Laboratory, California Institute of Technology under a contract with NASA.
GALEX is a NASA Small Explorer, launched in 2003 April, operated for NASA by the California Institute of Technology under NASA contract NAS5-98034.
The Pan-STARRS1 Surveys (PS1) have been made possible through contributions of the Institute for Astronomy, the University of Hawaii, the Pan-STARRS Project Office, the Max-Planck Society and its participating institutes, the Max Planck Institute for Astronomy, Heidelberg and the Max Planck Institute for Extraterrestrial Physics, Garching, The Johns Hopkins University, Durham University, the University of Edinburgh, Queen's University Belfast, the Harvard-Smithsonian Center for Astrophysics, the Las Cumbres Observatory Global Telescope Network Incorporated, the National Central University of Taiwan, the Space Telescope Science Institute, the National Aeronautics and Space Administration under Grant No. NNX08AR22G issued through the Planetary Science Division of the NASA Science Mission Directorate, the National Science Foundation under Grant No. AST-1238877, the University of Maryland, and Eotvos Lorand University (ELTE). The flowchart in this paper was created using the Miro online collaborative whiteboard platform (\url{www.miro.com}).
\end{acknowledgements}

\bibliographystyle{aa} 
\bibliography{refs.bib}

\begin{appendix}
\section{Clear and unclear merged \hi\ detections}
\label{p1:app:unclear_merged}

\begin{figure*}
        \centering
        \includegraphics[width = 1 \textwidth]{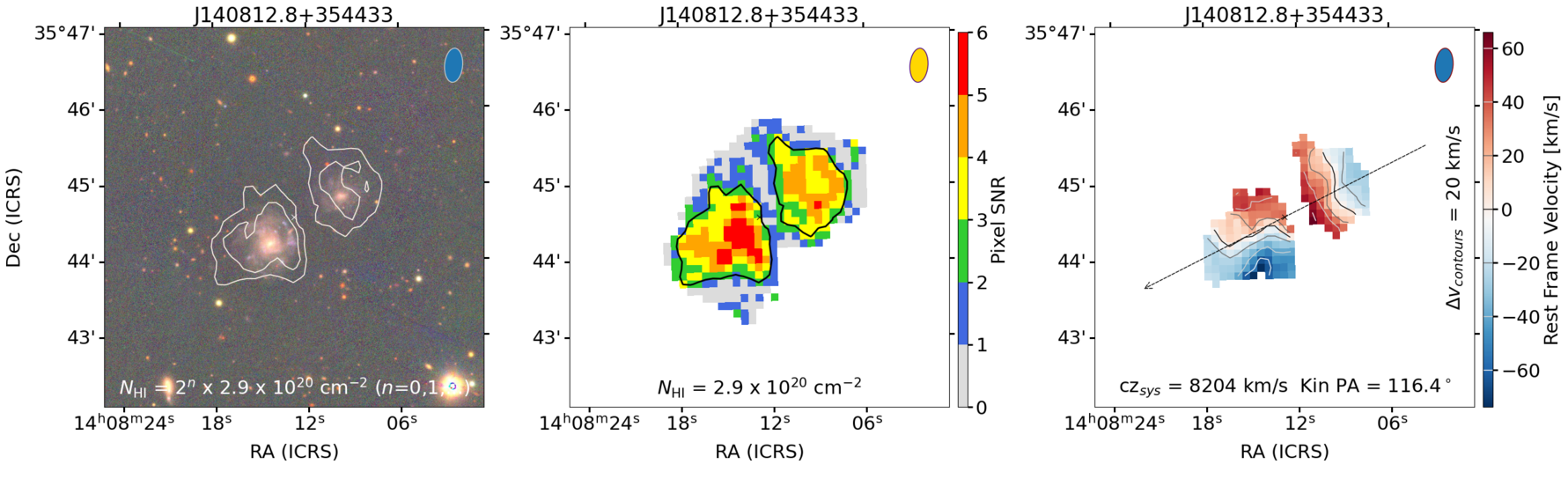}
        \caption{AHCJ140812.8+354433 as an example of a clear merged detection with two distinct optical counterparts within a well resolved \hi\ detection. \textit{Left:} PanSTARRS1 color image overlaid with \hi\ contours. \textit{Middle:} The S/N map of the \hi\ detection. \textit{Right:} Velocity field of the \hi\ detection. All panels are obtained from the output of SIP.}
        \label{p1:fig:J140812.82+354433.8}
    \end{figure*}

\begin{figure*}
        \centering
        \includegraphics[width = 1 \textwidth]{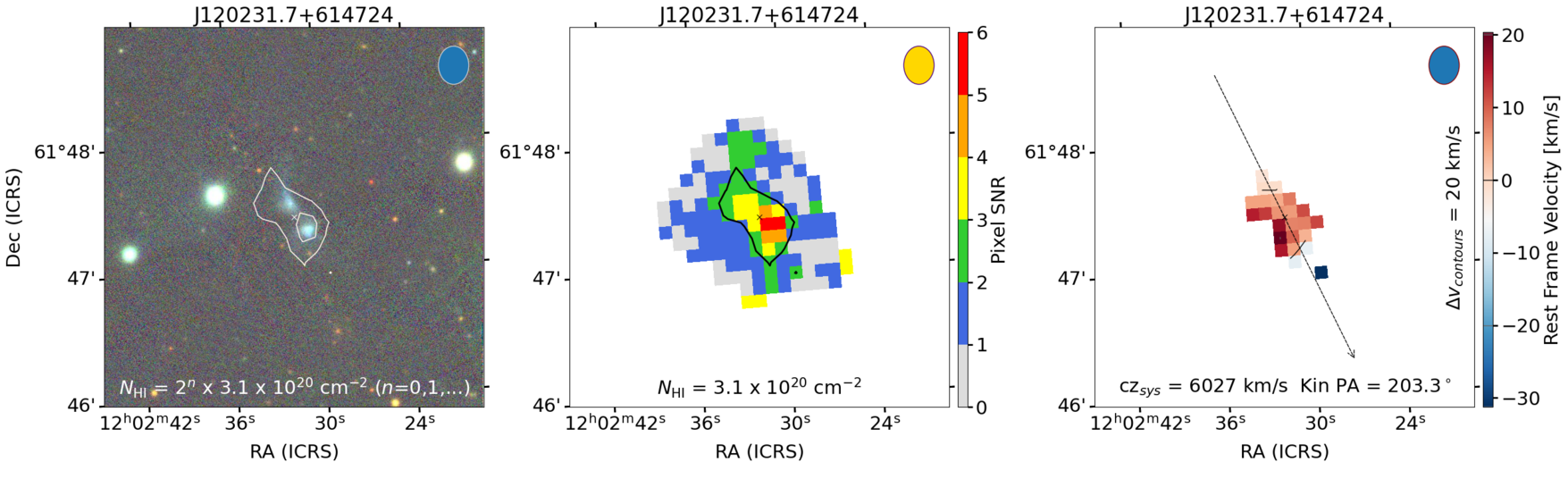}
        \caption{AHCJ120231.7+614724. Unclear merged detection with two distinct optical counterparts within a poorly resolved (with respect to the projected distance of optical counterparts) \hi\ detection. Panels are the same as in Fig. \ref{p1:fig:J140812.82+354433.8}.}
        \label{p1:fig:J120231.71+614724.8}
    \end{figure*}

\begin{figure*}
        \centering
        \includegraphics[width = 1 \textwidth]{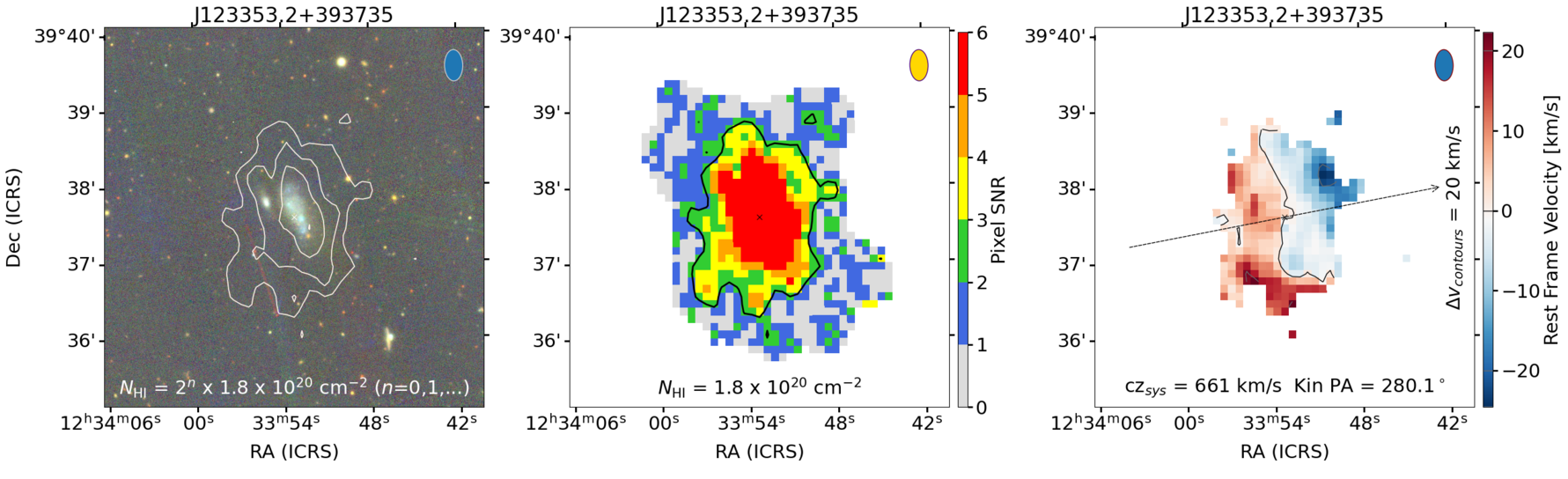}
        \caption{AHCJ123353.2+393735. Unclear merged detection with two distinct optical counterparts within a poorly resolved (with respect to the projected distance of optical counterparts) \hi\ detection. Panels are the same as in Fig. \ref{p1:fig:J140812.82+354433.8}.}
        \label{p1:fig:J123353.22+393735.3}
    \end{figure*}

\begin{figure*}
        \centering
        \includegraphics[width = 1 \textwidth]{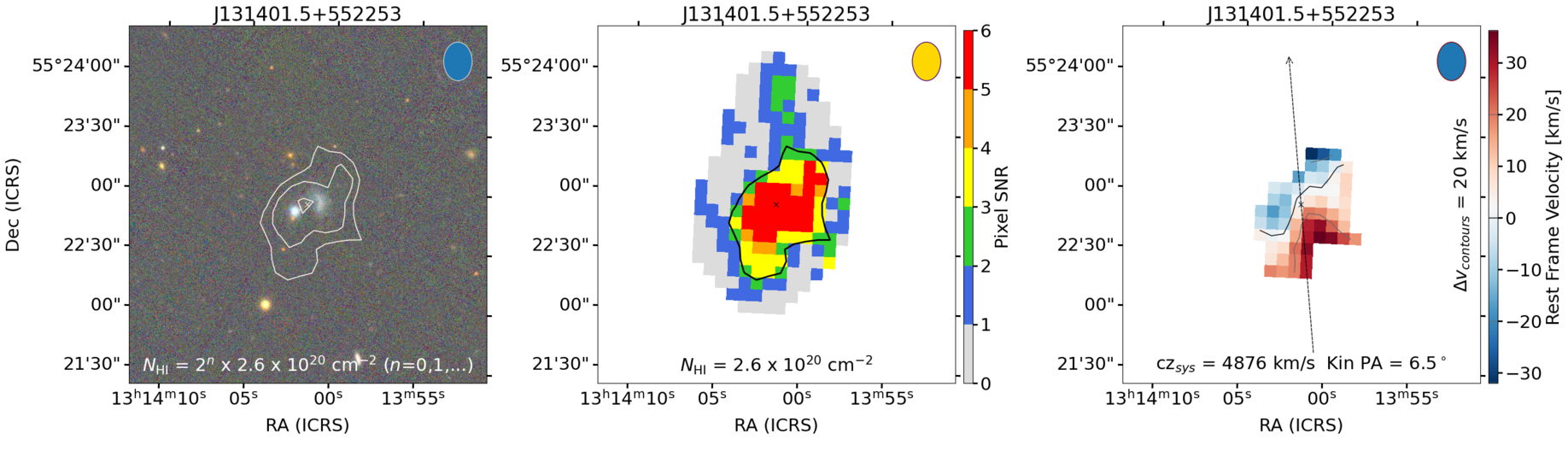}
        \caption{AHCJ131401.5+552253. Unclear merged detection with two distinct optical counterparts within a poorly resolved (with respect to the projected distance of optical counterparts) \hi\ detection. Panels are the same as in Fig. \ref{p1:fig:J140812.82+354433.8}.}
        \label{p1:fig:J131401.53+552253.4}
    \end{figure*}

\begin{figure*}
        \centering
        \includegraphics[width = 1 \textwidth]{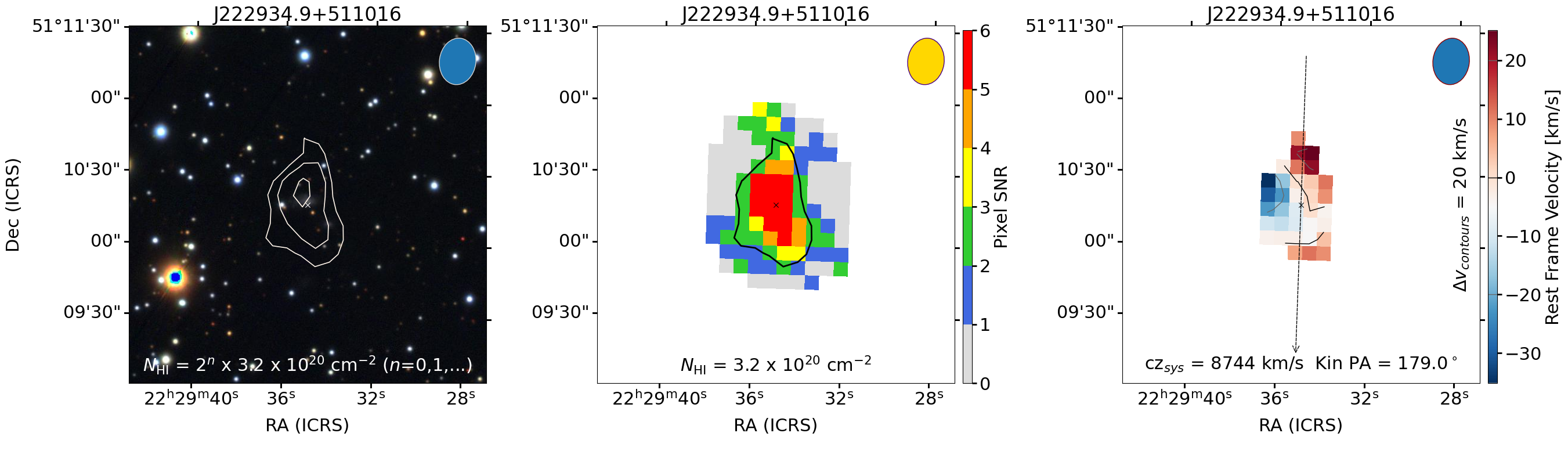}
        \caption{AHCJ222934.9+511016. Unclear merged detection with two distinct optical counterparts within a poorly resolved (with respect to the projected distance of optical counterparts) \hi\ detection. Panels are the same as in Fig. \ref{p1:fig:J140812.82+354433.8}.}
        \label{p1:fig:J222934.96+511016.8}
    \end{figure*}

\begin{figure*}
        \centering
        \includegraphics[width = 1 \textwidth]{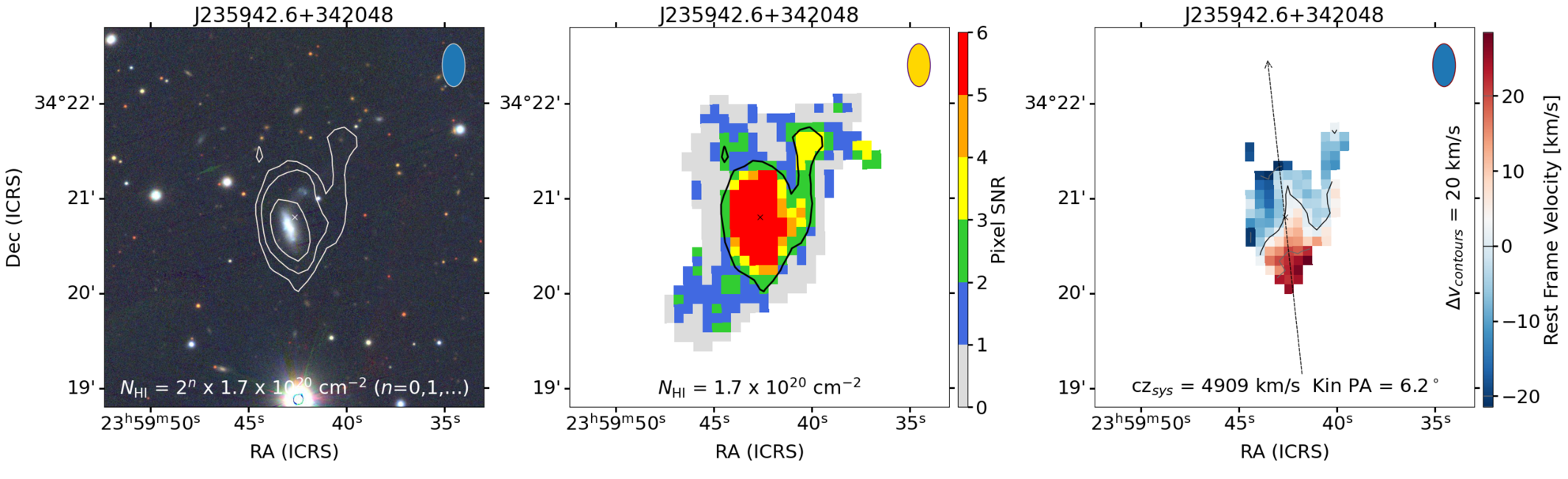}
        \caption{AHCJ235942.6+342048. Unclear merged detection with two distinct optical counterparts within a poorly resolved (with respect to the projected distance of optical counterparts) \hi\ detection. Panels are the same as in Fig. \ref{p1:fig:J140812.82+354433.8}.}
        \label{p1:fig:J235942.66+342048.2}
    \end{figure*}

\begin{figure*}
        \centering
        \includegraphics[width = 1 \textwidth]{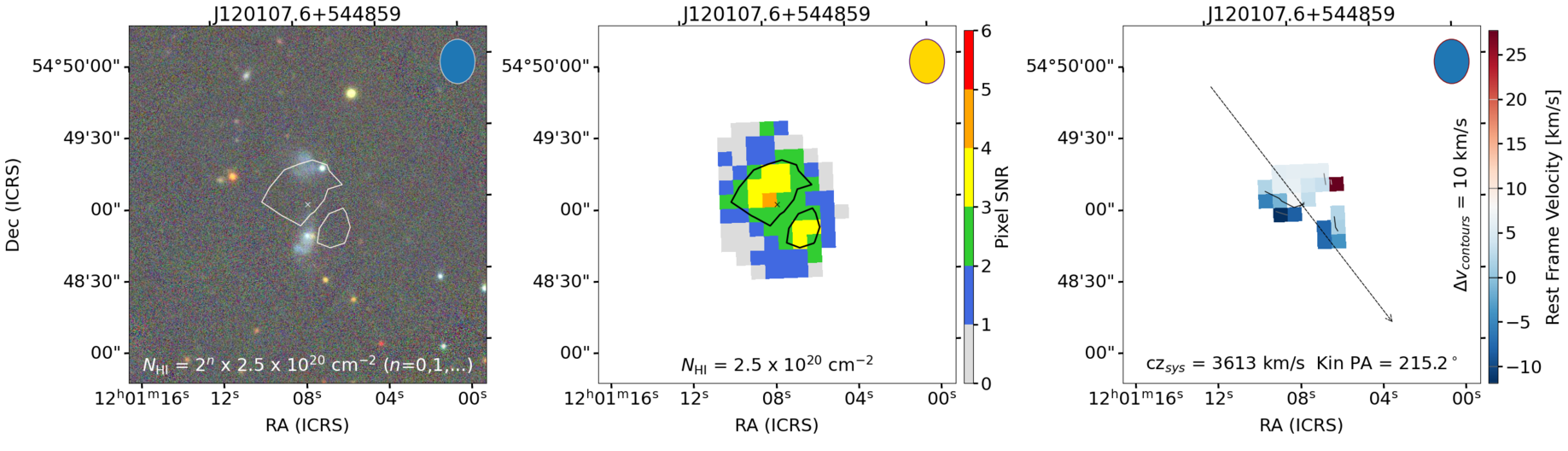}
        \caption{AHCJ120107.6+544859. Unclear merged detection with two distinct optical counterparts within a low S/N and poorly resolved (with respect to the projected distance of optical counterparts) \hi\ detection. Panels are the same as in Fig. \ref{p1:fig:J140812.82+354433.8}.}
        \label{p1:fig:J120107.60+544859.1}
    \end{figure*}

\begin{figure*}
        \centering
        \includegraphics[width = 1 \textwidth]{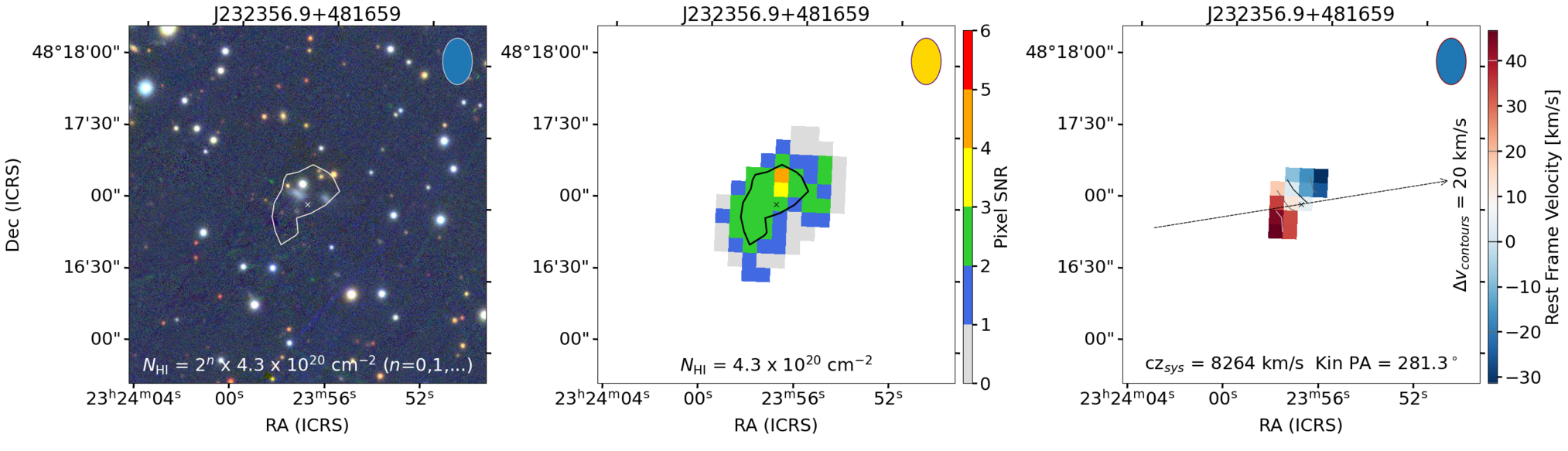}
        \caption{AHCJ232356.9+481659. Unclear merged detection with two distinct optical counterparts within a low S/N and poorly resolved (with respect to the projected distance of optical counterparts) \hi\ detection. Panels are the same as in Fig. \ref{p1:fig:J140812.82+354433.8}.}
        \label{p1:fig:J232356.99+481659.4}
    \end{figure*}

\begin{figure*}
        \centering
        \includegraphics[width = 1 \textwidth]{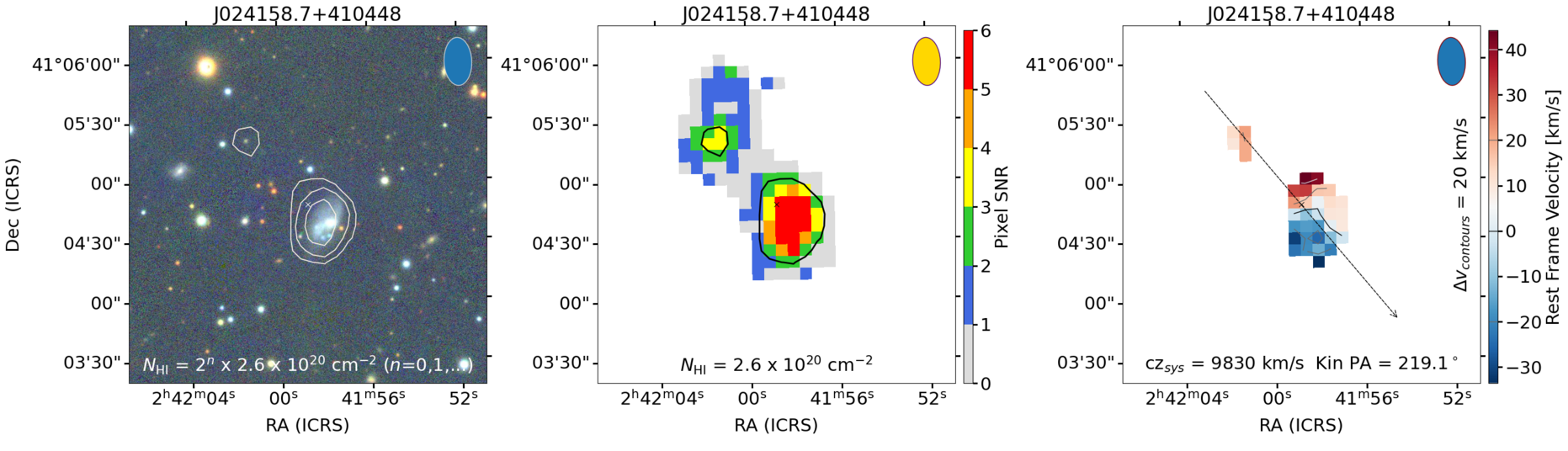}
        \caption{AHCJ024158.7+410448. Unclear merged detection with one clear optical counterpart and another (small) potential counterpart (in the northeast) within a low S/N \hi\ detection. Panels are the same as in Fig. \ref{p1:fig:J140812.82+354433.8}.}
        \label{p1:fig:J024158.76+410448.0}
    \end{figure*}

\begin{figure*}
        \centering
        \includegraphics[width = 1 \textwidth]{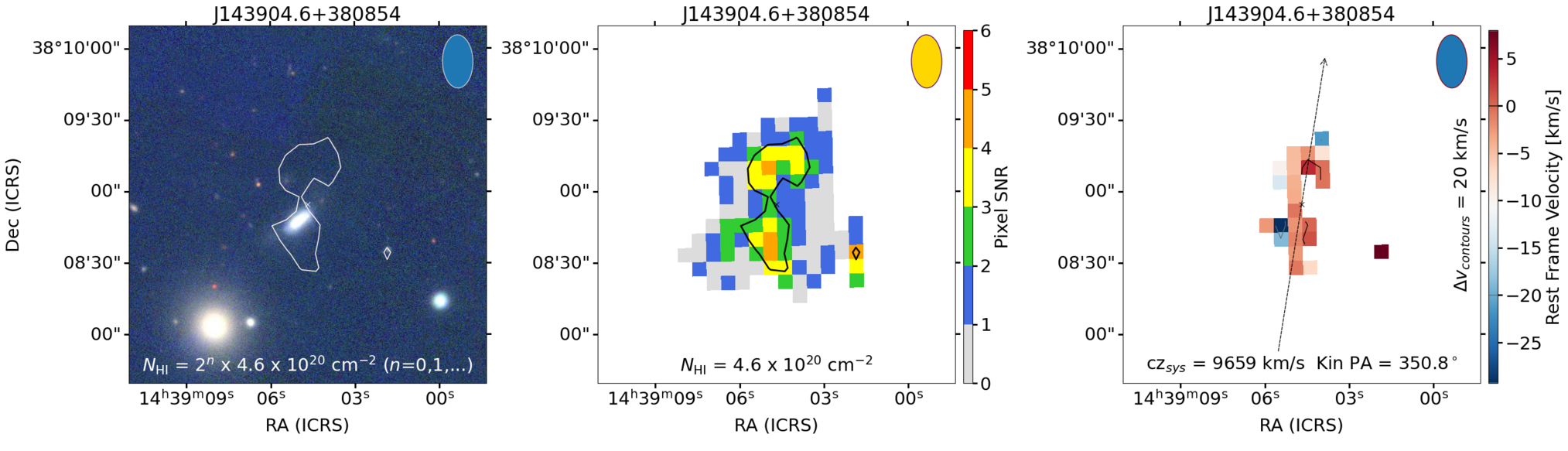}
        \caption{AHCJ143904.6+380854. Unclear merged detection with one clear optical counterpart and another (small) potential counterpart (in the southwest) within a low S/N \hi\ detection. Panels are the same as in Fig. \ref{p1:fig:J140812.82+354433.8}.}
        \label{p1:fig:J143904.68+380854.5}
    \end{figure*}

    In Fig. \ref{p1:fig:J140812.82+354433.8}, we show an example of a clear merged detection, where the two galaxies have clear optical counterparts and have distinct \hi\ emission. However, in some cases it was less clear if the \hi\ detection corresponded to the emission of one or more galaxies. We show these cases in Figs. \ref{p1:fig:J120231.71+614724.8} - \ref{p1:fig:J143904.68+380854.5}, and list them in Table \ref{p1:tab:unclear}. 
    In the following, we give a short description for each detection.

    \textbf{AHCJ120231.7+614724.} As seen in Fig. \ref{p1:fig:J120231.71+614724.8}, there are two optical counterparts inside the S/N > 3 \hi\ emission (left and middle panels). However, the emission is not resolved enough to clearly say if the \hi\ emission from the north-east galaxy is indeed detected or not. However, the velocity field in the right panel tentatively suggests different kinematics at the two locations.
    
    \textbf{AHCJ123353.2+393735.} The detection is shown in Fig. \ref{p1:fig:J123353.22+393735.3}. As in the example above, there are two optical counterparts in this detection, but the \hi\ emission cannot be distinguished between the two galaxies due to their close projected distances. However, 
    there is a clear increase in S/N at the position of the north-east galaxy (middle panel) accompanied by a local increase in velocity (right panel).

    \textbf{AHCJ131401.5+552253.} The detection is shown in Fig. \ref{p1:fig:J131401.53+552253.4}, where two optical counterparts are well inside the high S/N \hi\ emission (left and middle panels). As in the cases above, the emission is not resolved enough to clearly say if both, or only one of the galaxies are detected. However, the fact that the kinematic axis of the velocity field is misaligned with the morphological axis of the \hi\ detection indicates a potential detection of both galaxies.

    \textbf{AHCJ222934.9+511016.} The detection is shown in Fig. \ref{p1:fig:J222934.96+511016.8}. The reasoning is the same as the detection above; there are two likely optical counterparts contained within the high S/N \hi\ emission but not enough spatial resolution to distinguish if both optical galaxies are detected or not. Again, the velocity field offers additional evidence for a merged detection with anomalous velocity emission in the southwest extension where one of the two optical galaxies is located.

    \textbf{AHCJ235942.6+342048.} The detection is shown in Fig. \ref{p1:fig:J235942.66+342048.2}. As in the case of AHCJ123353.22+393735.3, there are two optical counterparts within the high S/N \hi\ emission accompanied by the distortion of the velocity field at the same location (north-west part of the S/N > 5 emission). The lower S/N < 4 emission in the north-west part of the detection is not considered to be coincident with a galaxy.

    \textbf{AHCJ120107.6+544859.} The detection is shown in Fig. \ref{p1:fig:J120107.60+544859.1}. As before, we see two clear optical counterparts to this emission, but the S/N of the \hi\ emission is too low and the resolution is insufficient to securely claim the detection of both galaxies.

    \textbf{AHCJ232356.9+481659.} The detection is shown in Fig. \ref{p1:fig:J232356.99+481659.4}. As in the previous galaxy, the S/N of the detection is low and the resolution is too insufficient to robustly claim detection of both optical galaxies.

    \textbf{AHCJ024158.7+410448.} The detection is shown in Fig. \ref{p1:fig:J024158.76+410448.0}. This time, the two optical counterparts are spatially distinct, but the detection of the north-east galaxy is unclear due to the low S/N of the detection combined with the very small optical counterpart, which could instead be a background galaxy coincident with a noise peak.

    \textbf{AHCJ143904.6+380854.} The detection is shown in Fig. \ref{p1:fig:J143904.68+380854.5}. As in the case above, there is the main \hi\ emission at the center with a clear optical counterpart, and another distinct small \hi\ emission peak (single pixel with S/N > 4) to the south-west. Due to the perfect spatial coincidence with a small optical counterpart, the south-west peak might indeed be an \hi\ detection, but the relatively low S/N of the detection combined with the very small spatial extent of detection, make it highly probable the counterpart is a background galaxy that happens to be coincident with a noise peak.

\begin{table}
\centering
\label{tab:general_properties}
\caption{Unclear merged detections.}
\begin{tabular}{lll}
\hline \hline
Apertif name & $N_{\rm sample}$ & reason \\
 AHC- & & \\
\hline 
J120231.7+614724 & 2 & res \\
J123353.2+393735 & 2 & res \\
J131401.5+552253 & 2 & res \\
J222934.9+511016 & 2 & res \\
J235942.6+342048 & 2 & res \\
J120107.6+544859 & 2 & res\&snr \\
J232356.9+481659 & 1 & res\&snr \\
J024158.7+410448 & 2 & snr\&uoc \\
J143904.6+380854 & 1 & snr\&uoc \\
\hline
\end{tabular}
\tablefoot{Apertif name is based on the position of the detection with the right ascension and declination in J2000 as: AHCJhhmmss.s+ddmmss. $N_{\rm sample}$ denotes the number of galaxies in the detection that entered our parent dwarf sample. All unclear detections are thought to contain at most two galaxies. The reason denotes why the detection was classified as unclear. Possibilities are: res - insufficient spatial resolution of the \hi\ emission, snr - too low S/N of the \hi\ detection, uoc - unclear optical counterpart, or a combination of these.}
\label{p1:tab:unclear}
\end{table}

\end{appendix}

\end{document}